%% file: BBC_version_2.tex
\documentclass[aps,prl,twocolumn,superscriptaddress,longbibliography]{revtex4-2}

\usepackage{amsmath,amssymb}
\usepackage{graphicx}
\usepackage{hyperref}
\usepackage{bm}
\usepackage{physics}
\usepackage{balance}

\hypersetup{
  colorlinks=true,
  citecolor=blue,
  linkcolor=blue,
  urlcolor=blue
}

\begin{document}

\title{Symmetry Protected Bulk-Boundary Correspondence in Interacting Topological Insulators}

\author{Kiran Babasaheb Estake}
\affiliation{Raman Research Institute, Bengaluru 560080, India}

\author{Dibyendu Roy}
\affiliation{Raman Research Institute, Bengaluru 560080, India}


\begin{abstract}
We establish a quantitative bulk–boundary correspondence in interacting topological insulators by relating many-body topological invariants to characteristic degeneracy structures in the entanglement spectrum. Focusing on generalized Su–Schrieffer–Heeger chains with higher winding number, we construct a gauge-invariant many-body winding invariant based on Pancharatnam geometric phases that remains well defined in the presence of interactions. We show that this invariant uniquely determines the low-lying entanglement-spectrum degeneracy, which exhibits a universal $4^{\nu}$ scaling with the winding number $\nu$, providing a concrete formulation of bulk–boundary correspondence beyond single-particle topology. Using exact diagonalization, we demonstrate the robustness of this correspondence under interactions and symmetry-preserving disorder, and identify inversion symmetry as a minimal protecting symmetry that stabilizes both the quantization of the invariant and the associated entanglement degeneracies. Our results unify geometric-phase invariants and entanglement diagnostics within a many-body framework and provide a route to identifying interacting topological phases beyond band theory.
\end{abstract}

\maketitle

\textit{Introduction—}
Topological phases of matter have emerged as a central paradigm in condensed matter physics, characterized by global topological invariants that are insensitive to local perturbations and give rise to robust boundary phenomena \cite{Schnyder2008, Kitaev2009}. A defining feature of these phases is the bulk–boundary correspondence (BBC), which relates quantized bulk topological invariants to the existence of protected edge or surface states. This principle is firmly established in noninteracting systems such as the integer quantum Hall effect and topological band insulators, where single-particle band topology predicts gapless boundary modes that are robust against symmetry-preserving perturbations \cite{Thouless1982, Hatsugai1993, Hasan2010}. Beyond its conceptual importance, the BBC provides a powerful framework for identifying and classifying topological phases through experimentally accessible boundary signatures \cite{Goldman2016, Klembt2018, Weber2022}.

While the BBC is well understood within the framework of noninteracting band theory, its extension to interacting systems remains a fundamental challenge \cite{Alisepahi2023, Sone2025}. Interactions can invalidate single-particle descriptions and alter topological classifications \cite{Chen2013, David2022, Li2023, Bisht2024}. 
Prominent examples include interaction-induced reductions of topological classifications and the interaction-driven gaping or removal of boundary zero modes that are protected in the noninteracting limit \cite{Fidkowski2010, Fidkowski2011, Turner2011, Gurarie2011}. As a result, bulk topological invariants defined in terms of single-particle Bloch bands are generally insufficient to characterize interacting phases, and a universally applicable formulation of BBC in interacting many-body systems beyond band theory remains less explicit.

To address the breakdown of band-theoretic invariants in interacting systems, several many-body diagnostics have been proposed \cite{Niu1984,  Wang2010, Wang2012, Manmana2012, Ke2017, Sone2024, Salvo2024}. In particular, the many-body Berry phase (MBBP) provides a natural generalization of geometric phases to interacting ground states and has been used to characterize polarization and topological properties beyond single-particle descriptions \cite{Ortiz1994, Resta1998, Grusdt2019, Le2020}. Independently, the entanglement spectrum (ES) has emerged as a powerful probe of topological order and symmetry-protected phases, with characteristic degeneracies reflecting nontrivial boundary physics even in the absence of physical edges \cite{Li2008, Pollmann2010, Turner2011}. Related connections between the ES and boundary physics or topological invariants in one-dimensional (1D) chiral insulators have been established in free-fermion settings \cite{Fidkowski2010(1), Monkman2023}. However, a quantitative BBC linking interacting many-body topological invariants to characteristic ES structures remains incomplete. 

In this Letter, we establish a quantitative symmetry-protected BBC in an interacting topological system by relating many-body topological invariants to characteristic degeneracies of the ES, which act as boundary proxies associated with the virtual edges created by an entanglement bipartition. Focusing on the Su–Schrieffer–Heeger (SSH) model with chiral symmetry-preserving density–density interactions \cite{Su1979}, we first analyze the noninteracting limit, where phases distinguished by a quantized Berry phase exhibit distinct ES degeneracies, and then demonstrate numerically that this correspondence persists in the presence of interactions. Extending the model to higher winding numbers, we show that the Berry phase, being defined modulo $2\pi$, cannot distinguish phases differing by multiples of $2\pi$. We therefore introduce a many-body winding invariant constructed from gauge-invariant Pancharatnam overlaps \cite{Pancharatnam1956, Samuel1988}, which remains quantized in the interacting systems and resolves phases with winding numbers $\nu=0,1,2$. We demonstrate that this invariant dictates the low-lying ES structure, enforcing a characteristic degeneracy scaling $4^\nu$. Finally, we establish that inversion symmetry alone protects both the invariant and the ES degeneracies against disorder, thereby preserving the interacting BBC in this model, even in the absence of chiral symmetry. We further show that the above findings also hold in synthetic higher dimensions.

\begin{figure}[t]
  \centering
  \includegraphics[width=\columnwidth]{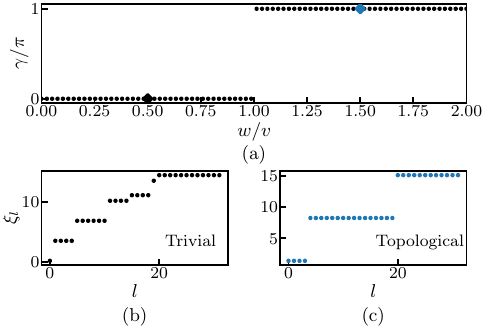}
  \caption{BBC in the noninteracting SSH model.
(a) MBBP $\gamma$ (in units of $\pi$) vs $w/v$. Black and blue markers denote representative trivial ($w/v=0.5$) and topological ($w/v=1.5$) points.
(b) Lowest 32 entanglement levels $\xi_l$ in the trivial phase.
(c) Lowest 32 entanglement levels $\xi_l$ in the topological phase, showing the characteristic fourfold degeneracy. All plots are generated for $L=10$ unit cells, i.e., $20$ sites.}
  \label{fig1:free_ssh_bbc}
\end{figure}

\textit{Noninteracting BBC—}We consider the noninteracting SSH model \cite{Su1979} at half filling, described by the Hamiltonian
\begin{align}
	\hat{H}_0=\sum_j \Bigl(v\hat{c}_{j,A}^\dagger \hat{c}_{j,B}+w\hat{c}_{j+1,A}^\dagger \hat{c}_{j,B}+\text{H.c.}\Bigr),  
\end{align}
where $\hat{c}_{j,A/B}^\dagger$ and $\hat{c}_{j,A/B}$ are the fermionic creation and annihilation operators at unit cell $j$ and sublattice $A/B$. $\hat{H}_0$ possesses time-reversal, particle-hole, and chiral symmetries as internal symmetries, placing it in symmetry class BDI. It also has spatial inversion symmetry. To characterize the bulk topology at the many-body level, we impose a twisted boundary condition and compute the MBBP ($\gamma$) accumulated by the many-body ground state under adiabatic insertion of a $2\pi$ flux. In the thermodynamic limit, $\gamma$ is exactly quantized to $0$ for $v>w$ and $\pi$ for $v<w$, as follows from an analytical derivation, with a topological transition at $v=w$ (Fig. \ref{fig1:free_ssh_bbc}(a)) \cite{SuppMat2026}.

To probe the corresponding boundary physics without introducing physical edges, we compute the ES. It consists of entanglement levels $\xi_l$ obtained by bipartitioning the periodic chain and captures the spectrum of virtual edge degrees of freedom at the entanglement cut. Figs. \ref{fig1:free_ssh_bbc}(b) and (c) show the sorted ES for representative points in the trivial and topological phases. In the trivial phase ($\gamma=0$), the entanglement levels (especially the lowest level) are generically nondegenerate. In contrast, in the topological phase ($\gamma=\pi$) the ES exhibits a characteristic global fourfold degeneracy, which can be understood analytically in terms of the zero-mode structure of the entanglement Hamiltonian \cite{SuppMat2026}. 

Taken together, the quantized MBBP and the associated ES degeneracies provide an explicit demonstration of BBC in the noninteracting SSH model at the many-body level, which will serve as the reference point for our analysis of interacting, higher-winding, and disordered extensions of the model.

\textit{Interacting BBC—}
We now introduce interactions and examine the fate of the BBC beyond band theory. Specifically, we consider density–density interactions within and between unit cells,
\begin{align}
	\hat{H}_\text{int} &= \frac{U}{2} \sum_j \Bigl(\hat{n}_{j, A}-\frac{1}{2}\Bigr) \Bigl(\hat{n}_{j, B}-\frac{1}{2}\Bigr) \notag\\
	&+ \frac{V}{2} \sum_j \Bigl(\hat{n}_{j+1, A}-\frac{1}{2}\Bigr) \Bigl(\hat{n}_{j, B}-\frac{1}{2}\Bigr),
\end{align}
where $\hat{n}_{j, A/B}=\hat{c}_{j, A/B}^\dagger \hat{c}_{j, A/B}$ is the number operator. Now the full Hamiltonian is $\hat{H}_I=\hat{H}_0+\hat{H}_{\text{int}}$. Written in this form, these interactions preserve the internal and spatial symmetries of $\hat{H}_0$, including chiral and inversion symmetries. Nevertheless, due to the presence of interactions the single-particle band description is no longer applicable for $\hat{H}_I$, and the persistence of the BBC therefore becomes a nontrivial many-body question.
\begin{figure}[t]
	\centering
	\includegraphics[width=\columnwidth]{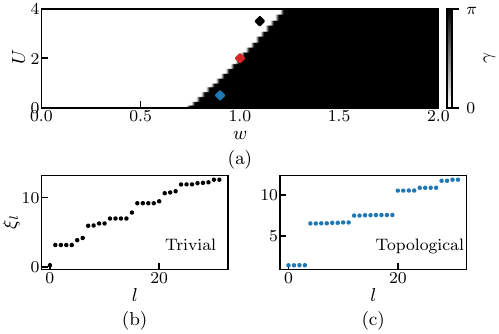}
	\caption{Interacting BBC.
		(a) MBBP $\gamma$  of the interacting SSH model as a function $w$ and $U$ for fixed $v=1$ and $V=2$. Interactions shift the topological phase boundary. This plot is generated for $L=10$.
		(b) Lowest 32 entanglement levels $\xi_l$ at $U=3.5$ and $w=1.1$ (black diamond in (a)), which lies in the trivial phase despite being topological in the noninteracting limit.
		(c) Lowest 32 entanglement levels at $U=0.5$ and $w=0.9$ (blue diamond in (a)), which lies in the topological phase despite being trivial in the noninteracting limit, exhibiting a fourfold degeneracy. (b) and (c) are generated for $L=12$.}
	\label{fig2:int_ssh_bbc}
\end{figure}

We compute the MBBP numerically using twisted boundary conditions. Fig. \ref{fig2:int_ssh_bbc}(a) shows that the Berry phase remains sharply quantized to $\gamma=0, \pi$ in the presence of interactions. For $U=V$, the topological transition remains pinned at $v=w$ (red diamond in Fig. \ref{fig2:int_ssh_bbc}(a)). In contrast, when $U\neq V$, the transition point shifts toward larger or smaller values of $w$, depending on the relative strength of intra- and intercell interactions. This interaction-induced shift is well captured by a mean-field renormalization of the effective hopping amplitudes \cite{SuppMat2026}. As a consequence, interactions can invert the topological character of specific parameter points, rendering phases that are trivial in the noninteracting limit topological, and vice versa, as illustrated by the representative black and blue diamonds in Fig. \ref{fig2:int_ssh_bbc}(a). Similar interaction-driven topology has been explored in nonlinear frameworks via self-consistent descriptions and nonlinear Chern numbers \cite{Sone2024}.

\begin{figure}[t]
	\centering
	\includegraphics[width=\columnwidth]{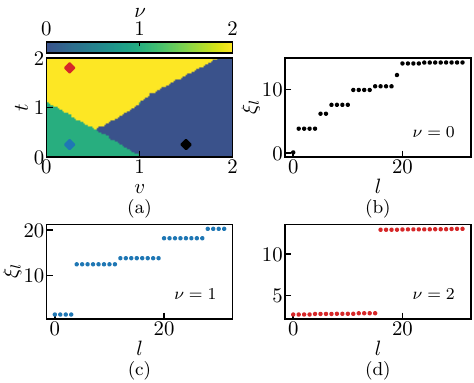}
	\caption{Many-body winding invariant and higher-winding BBC.
(a) Phase diagram of the interacting generalized SSH model showing the many-body winding invariant $\nu$ as a function of $v$ and $t$, for fixed $w=1$ and $U=V=1$ and for $L=10$. Lowest 32 entanglement levels $\xi_l$ at a representative point in the trivial phase ($\nu=0$) in (b), in the topological phase with $\nu=1$, exhibiting a characteristic fourfold degeneracy in (c), and in the topological phase with $\nu=2$, exhibiting a sixteenfold degeneracy in (d), consistent with the scaling $4^\nu$.
          (b), (c) and (d) are computed for $L=12$.}
	\label{fig3:int_gen_ssh_bbc}
\end{figure}
To probe the corresponding boundary physics, we compute the ES of the interacting ground state. As shown in Fig. \ref{fig2:int_ssh_bbc}(b), the ES remains generically nondegenerate in the trivial phase, while in Fig. \ref{fig2:int_ssh_bbc}(c) for the topological phase, the ES exhibits a characteristic fourfold degeneracy in the low-lying levels. Importantly, the onset of this degeneracy tracks precisely the interaction-shifted transition point extracted from the MBBP.

These results demonstrate that the BBC persists in the interacting regime, with quantized MBBP and corresponding low-lying ES degeneracies.

\textit{Higher winding beyond band topology—}
To access topological phases beyond the trivial ($\gamma=0$) and topological ($\gamma=\pi$) phases, we generalize the SSH model by including long-range hopping \cite{Chen2020},
\begin{align}
\hat{H}_t=t\sum_j (\hat{c}_{j+2,A}^\dagger \hat{c}_{j,B}+ \hat{c}_{j,B}^\dagger \hat{c}_{j+2,A}),
\end{align}
such that the full Hamiltonian is now $\hat{H}_{gen}=\hat{H}_0+\hat{H}_t+\hat{H}_{\text{int}}$. This generalized model also has chiral symmetry and it supports phases with higher winding numbers ($\nu$) in the noninteracting limit. While the MBBP provides a robust topological diagnostic in both noninteracting and interacting systems, its ability to resolve topological structure is fundamentally limited by its definition modulo $2\pi$. In the standard SSH model, this limitation is inconsequential because only two phases occur, distinguished by MBBP, $\gamma=0$ and $\pi$. However, when longer-range hopping is included, the noninteracting model supports phases with higher winding numbers. In this case, the MBBP alone cannot distinguish phases whose winding numbers differ by an even integer, assigning identical values to phases with $\nu=0$ and $\nu=2$. In noninteracting systems, this distinction is captured by an integer-valued single-particle winding number defined using Bloch vector associated Bloch Hamiltonian. In the presence of interactions, however, such a band-theoretic invariant is no longer well defined.
To overcome this limitation, we introduce a many-body winding invariant that generalizes the notion of winding beyond band theory and remains well defined in interacting systems.

We construct the many-body winding invariant by tracking the relative phase structure of the interacting ground state along the flux-insertion path with respect to a fixed reference state. 
Consider the family of ground states $\ket{G(\phi)}$ obtained under adiabatic insertion of a boundary twist $\phi$, and fix $\ket{G(0)}$ as a reference. 
For successive values $\phi_j$ and $\phi_{j+1}$, we evaluate the Pancharatnam triple overlap
\[
\langle G(0)|G(\phi_j)\rangle 
\langle G(\phi_j)|G(\phi_{j+1})\rangle
\langle G(\phi_{j+1})|G(0)\rangle.
\]
The argument of this quantity measures the local relative phase twist of the ground-state path with respect to the reference state. Summing these contributions over a full $2\pi$ flux cycle yields an accumulated phase winding. 
Dividing its total magnitude by $\pi$ defines a many-body winding invariant $\nu$.

\textit{Higher-winding BBC—}
We now apply this many-body winding invariant to the interacting generalized SSH model with longer-range hopping. Fig. \ref{fig3:int_gen_ssh_bbc}(a) shows the resulting phase diagram of the invariant $\nu$ as a function of $v$ and $t$, for fixed $w=1$ and $U=V=1$. Three distinct phases with winding numbers $\nu=0,1,$ and $2$ are clearly resolved. Notably, phases with $\nu=0$ and $\nu=2$, which are indistinguishable at the level of the MBBP, are unambiguously separated by the winding invariant. The sharp phase boundaries and integer quantization demonstrate that the invariant remains well defined in the presence of interactions.

To establish the corresponding boundary physics, we compute the ES at representative points from each phase, shown in Figs. \ref{fig3:int_gen_ssh_bbc}(b–d). In the trivial phase ($\nu=0$), the ES, especially the lowest entanglement level is nondegenerate (Fig. \ref{fig3:int_gen_ssh_bbc}(b)). For $\nu=1$, the low-lying ES exhibits a characteristic fourfold degeneracy (Fig. \ref{fig3:int_gen_ssh_bbc}(c)), consistent with the interacting SSH phase discussed above. Strikingly, in the $\nu=2$ phase the ES displays a sixteenfold degeneracy in the low-lying levels (Fig. \ref{fig3:int_gen_ssh_bbc}(d)). We thus uncover a universal scaling relation between bulk topology and entanglement structure: the low-lying ES degeneracy scales as $4^\nu$ \cite{ES_Degeneracies_Comment}. In the interacting case, higher entanglement levels are generically split, but the low-lying multiplets remain protected. The resulting one-to-one correspondence between the integer-valued bulk invariant $\nu$ and the degeneracy structure of the low-lying ES establishes a higher-winding many-body BBC beyond the reach of the MBBP.

\begin{figure}[t]
	\centering
	\includegraphics[width=\columnwidth]{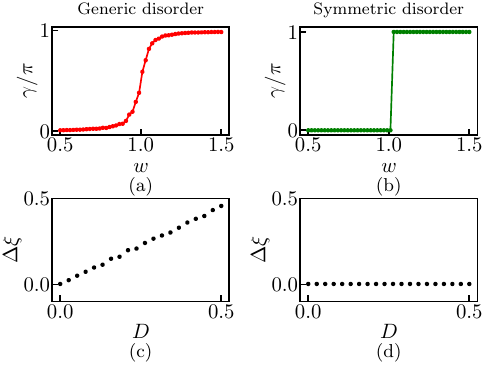}
	\caption{Symmetry-protected many-body topology and BBC in the presence of disorder. 
Generic disorder: 
(a) MBBP $\gamma$ as vs $w$ for $v=1$, $U=V=1$ and $D=0.7$, showing a smooth crossover between $\gamma=0$ and $\pi$. 
(c) Splitting $\Delta \xi=\xi_4-\xi_1$ of the lowest entanglement levels as a function of $D$ in the topological phase ($v=0.5$, $w=1$), depicting lifting of the characteristic degeneracy. Inversion-symmetric disorder: 
(b) MBBP for $D=0.7$, where a sharp topological transition is restored. 
(d) $\Delta \xi$ in the topological phase ($v=0.5$, $w=1$), where the degeneracies remain exact, signaling symmetry-protected BBC. Both $\gamma$ and $\Delta \xi$ are calculated by averaging over $300$ disorder realizations for $L=7$.}
	\label{fig4:int_dis_ssh_bbc}
\end{figure}

\textit{Symmetry protection and disorder—}
We now test the symmetry-protected nature of the many-body BBC by introducing onsite disorder into the interacting SSH model: 
\begin{align}
\hat{H}_{\mathrm{dis}}=\sum_j\left(\epsilon_{j,A}\hat{n}_{j,A}+\epsilon_{j,B} \hat{n}_{j,B}\right),
\end{align}
such that the full Hamiltonian is now $\hat{H}_{D}=\hat{H}_0+\hat{H}_{\text{int}}+\hat{H}_{\mathrm{dis}}$. The onsite energies $\epsilon_{j,\alpha}$ with $\alpha = A,B$, are independently drawn from a uniform distribution $\epsilon_{j,\alpha} \in [-D, D]$, with disorder strength $D$. 

We consider two cases: generic disorder, which breaks translation, chiral, and inversion symmetries at the level of individual realizations, and inversion-symmetric disorder, constructed such that inversion symmetry is preserved but chiral symmetry remains broken in every realization \cite{SuppMat2026}.

We first examine the behavior of the MBBP. Fig. \ref{fig4:int_dis_ssh_bbc} shows $\gamma$ as a function of $w$ for fixed $v=1$ in the presence of interactions $U=V=1$. In the case of generic disorder, the MBBP changes smoothly from $\gamma=0$ to $\pi$ in Fig. \ref{fig4:int_dis_ssh_bbc}(a), indicating that the sharp topological transition observed in the clean system is replaced by a crossover. In contrast, when disorder preserves inversion symmetry, the MBBP remains sharply quantized and exhibits a well-defined transition point (Fig. \ref{fig4:int_dis_ssh_bbc}(b)), revealing that inversion symmetry alone is sufficient to protect the many-body topology even in the absence of chiral symmetry.

The many-body winding invariant exhibits the same symmetry-dependent behavior. For generic disorder, the invariant loses quantization and fluctuates strongly. When inversion symmetry is enforced, however, the winding invariant remains sharply quantized and continues to reproduce the clean-system phase structure \cite{SuppMat2026}.

To probe the corresponding boundary physics, we analyze the ES of the interacting ground state. Rather than plotting the full spectrum, we focus on the splitting $\Delta \xi=\xi_4-\xi_1$ between the lowest entanglement levels, which provides a sensitive diagnostic of degeneracy lifting. As shown in Fig. \ref{fig4:int_dis_ssh_bbc}(c), generic disorder lifts the characteristic low-energy fourfold degeneracy in the topological phase as disorder strength $D$ is increased, resulting in a finite $\Delta \xi$. This reflects the breakdown of the BBC once protecting symmetries are removed.

Strikingly, when inversion symmetry is preserved, the low-energy ES levels remain degenerate in the topological phase. Fig. \ref{fig4:int_dis_ssh_bbc}(d) shows that $\Delta \xi$ remains pinned to zero in the presence of inversion-symmetric disorder, directly mirroring the quantization of the bulk invariants. This demonstrates that the many-body BBC persists in the presence of disorder provided inversion symmetry is maintained.

Together, these results establish that inversion symmetry alone is sufficient to stabilize the interacting BBC even in absence of chiral symmetry. The role of inversion symmetry as a  protector of 1D topology, independent of chiral symmetry, has been emphasized in \cite{Fuchs2021}, where the Zak phase and polarization were shown to depend sensitively on orbital embedding. Our results extend this perspective to interacting systems, where inversion symmetry likewise protects both the many-body invariant and the associated ES structure.

Finally, we note that the entanglement-based BBC developed here extends naturally to systems that realize higher-dimensional topology through synthetic parameters. In particular, an SSH chain with a periodically modulated onsite potential provides a realization of a Thouless pump, where the pump parameter plays the role of a synthetic momentum, mapping the system to an effective two-dimensional Chern insulator \cite{Li2014}. The Hamiltonian for this model with interaction is given by $\hat{H}_{IT}=\hat{H}_{0}+\hat{H}_{\mu}+\hat{H}_{\text{int}}$, where 
\begin{align}
\hat{H}_{\mu}=\sum_j \Bigl(\mu_A\hat{c}_{j,A}^\dagger \hat{c}_{j,A}+\mu_B\hat{c}_{j,B}^\dagger \hat{c}_{j,B}\Bigr),
\end{align}
and the parameters are parameterized as $v=1+\delta \cos \theta,$
$w=1-\delta \cos \theta,$
$\mu_A=g_A+h\cos(\theta-\tilde{\phi})$ and
$\mu_B=g_B+h\cos(\theta+\tilde{\phi}).$ We keep $U=V$ in this model.
Our framework captures this extension in both the noninteracting and interacting regimes \cite{SuppMat2026}: the correlation-matrix spectrum exhibits spectral flow associated with the Chern number ($C$), while the many-body ES flow displays characteristic degeneracy structures that follow the $4^{|C|}$ scaling. Moreover, the many-body winding invariant introduced in this work directly reproduces the Chern number through its evolution over the pump cycle. These results demonstrate that our approach is not restricted to strictly 1D systems, but provides a unified description of BBC across dimensions, including interacting topological pumps.

\textit{Summary and outlook—}
We established a symmetry-protected BBC in interacting systems by relating quantized many-body topological invariants to characteristic degeneracy structures in the ES. Going beyond the MBBP, we introduced a gauge-invariant, Pancharatnam-based winding invariant that resolves higher winding numbers and remains quantized in the presence of interactions, with the corresponding phases exhibiting a universal $4^{\nu}$ scaling of low-lying ES degeneracies. We further demonstrated that inversion symmetry constitutes a minimal protecting symmetry, stabilizing both the quantization of the invariant and the associated ES even in the absence of chiral symmetry. 

Our results establish a general framework for diagnosing interacting topological phases using purely many-body quantities and open several directions for future work. These include extensions to higher-dimensional and synthetic systems, with nontrivial Chern topology, as well as to non-Hermitian settings, where the interplay of interactions, symmetry, and topology remains largely unexplored. More broadly, our work provides a route toward a unified understanding of BBC beyond band theory.

\textit{Acknowledgments—} We thank Vijay Shenoy and Soumya Bera for useful discussions.


\clearpage
\onecolumngrid
\setcounter{NAT@ctr}{0}   
\input{SM}

\end{document}

%% file: SM.tex




\section*{Supplementary Material for: \\
	Symmetry Protected Bulk-Boundary Correspondence in Interacting Topological Insulators}

\setcounter{secnumdepth}{3}

\renewcommand{\thesection}{S\arabic{section}}
\renewcommand{\thesubsection}{S\arabic{section}.\arabic{subsection}}

\renewcommand{\thefigure}{S\arabic{figure}}
\renewcommand{\thetable}{S\arabic{table}}
\renewcommand{\theequation}{S\arabic{equation}}
\setcounter{section}{0}
\renewcommand{\thesection}{S\arabic{section}}
\renewcommand{\thesubsection}{S\arabic{section}.\arabic{subsection}}
\renewcommand{\thefigure}{S\arabic{figure}}
\renewcommand{\thetable}{S\arabic{table}}
\renewcommand{\theequation}{S\arabic{equation}}

\section{Noninteracting SSH Model: Analytical Details}\label{ES_non_int}

\subsection{Derivation of the many-body Berry phase for noninteracting fermions}

In this section, we derive analytically the many-body Berry phase (MBBP) of the half-filled noninteracting SSH model \cite{Su1979, Ortiz1994, Resta1998}. 
Although in the main text the MBBP is computed numerically in a gauge, where the Hamiltonian is strictly periodic in the flux $\phi$, that gauge breaks translational invariance.  
For an analytical treatment, it is more convenient to work in a transitionally invariant gauge, in which the flux is distributed uniformly along the chain.

We therefore consider the Hamiltonian
\begin{align}
\hat{H}(\phi)=
v\sum_{j=1}^{L} c_{j,A}^\dagger c_{j,B}
+w\,e^{-i\phi/L}\sum_{j=1}^{L-1} c_{j+1,A}^\dagger c_{j,B}
+{\rm H.c.},
\label{eq:uniform_gauge}
\end{align}
which describes the same physical system as the boundary-flux gauge in the main text, but it preserves lattice translation symmetry.  

Assuming periodic boundary condition and introducing Fourier-transformed operators,
\begin{equation}
c_{j,\alpha}=\frac{1}{\sqrt{L}}\sum_{k}e^{ijk}c_{k,\alpha},
\end{equation}
the Hamiltonian becomes
\begin{align}
\hat{H}(\phi)
=\sum_k 
\begin{pmatrix} 
c_{k,A}^\dagger & c_{k,B}^\dagger 
\end{pmatrix}
\begin{pmatrix}
0 & v+we^{-i(k+\phi/L)} \\
v+we^{i(k+\phi/L)} & 0
\end{pmatrix}
\begin{pmatrix} 
c_{k,A} \\ c_{k,B}
\end{pmatrix}.
\end{align}
The flux therefore enters only through the shifted momentum
\(
\kappa = k+\phi/L .
\)
As $\phi$ is increased from $0$ to $2\pi$, all momenta are translated by one discrete step $2\pi/L$ within the Brillouin zone.
The lower-band creation operator is
\begin{equation}
\eta_{\kappa,-}^\dagger
=-\frac{1}{\sqrt{2}}e^{i\theta(\kappa)}c_{k,A}^\dagger
+\frac{1}{\sqrt{2}}c_{k,B}^\dagger ,
\end{equation}
where $\theta(\kappa)=\arg\!\left[v+we^{-i\kappa}\right]$.
At half filling, the many-body ground state is obtained by filling all $L$ lower-band states,
\begin{equation}
\ket{G(\phi)}=\prod_{\kappa=\phi/L}^{2\pi+\phi/L} 
\eta_{\kappa,-}^\dagger\ket{0}.
\label{eq:GSphi}
\end{equation}

Because the Hamiltonian in this transitionally invariant gauge is not periodic in $\phi$, i.e., $\hat{H}(0)\neq \hat{H}(2\pi)$, the MBBP takes the form
\begin{equation}
\gamma = 
i\!\int_{0}^{2\pi}\! d\phi\, 
\bra{G(\phi)}\frac{d}{d\phi}\ket{G(\phi)}
+ \mathrm{Arg}\,\braket{G(0)}{G(2\pi)} .
\label{eq:berry_general}
\end{equation}

Differentiating Eq.~(\ref{eq:GSphi}) and taking the inner product with $\bra{G(\phi)}$ yields
\begin{equation}
\bra{G(\phi)}\frac{d}{d\phi}\ket{G(\phi)}
= \sum_{\kappa=\phi/L}^{2\pi+\phi/L}
\bra{0}\eta_{\kappa,-}
\frac{d}{d\phi}\eta_{\kappa,-}^\dagger
\ket{0}.
\end{equation}
In the thermodynamic limit, the sum becomes an integral,
\begin{equation}
\bra{G(\phi)}\frac{d}{d\phi}\ket{G(\phi)}
=\frac{L}{2\pi}\!\int_{\phi/L}^{2\pi+\phi/L}\!
d\kappa\,
\bra{0}\eta_{\kappa,-}
\frac{d}{d\phi}\eta_{\kappa,-}^\dagger
\ket{0}.
\end{equation}
Using $d/d\phi=(1/L)d/d\kappa$ gives
\begin{equation}
\bra{G(\phi)}\frac{d}{d\phi}\ket{G(\phi)}
=\frac{1}{2\pi}\!\int_{\phi/L}^{2\pi+\phi/L}\!
d\kappa\,
\bra{0}\eta_{\kappa,-}
\frac{d}{d\kappa}\eta_{\kappa,-}^\dagger
\ket{0}.
\label{eq:GdG}
\end{equation}

A straightforward evaluation gives
\begin{equation}
\bra{0}\eta_{\kappa,-}
\frac{d}{d\kappa}\eta_{\kappa,-}^\dagger
\ket{0}
= \frac{i}{2}\frac{d\theta(\kappa)}{d\kappa}
+\frac{i(L+1)}{2}.
\end{equation}
The second term is an extensive phase originating from the $k$-dependence of the Fourier operators.

Substituting into Eq.~(\ref{eq:GdG}) and integrating over $\phi$ yields
\begin{equation}
i\!\int_{0}^{2\pi}\! d\phi\,
\bra{G(\phi)}\frac{d}{d\phi}\ket{G(\phi)}
= -\frac{1}{4\pi}
\!\int_{0}^{2\pi}\! d\phi
\!\int_{\phi/L}^{2\pi+\phi/L}\!
d\kappa\,\frac{d\theta}{d\kappa}
-\pi(L+1).
\label{eq:int_term}
\end{equation}

The overlap between the initial and final ground states is
\begin{align}
	\bra{G(0)}\ket{G(2\pi)}&=\bra{0}\eta_{2\pi-2\pi/L, -} \eta_{2\pi-4\pi/L, -}...\eta_{4\pi/L, -} \eta_{2\pi/L, -} \eta_{0, -}\eta_{2\pi/L, -}^\dagger \eta_{4\pi/L, -}^\dagger ...\eta_{2\pi-2\pi/L, -}^\dagger \eta_{0, -}^\dagger \ket{0}\notag\\
	&=(-1)^{L-1}\bra{0}\eta_{2\pi-2\pi/L, -} \eta_{2\pi-4\pi/L, -}...\eta_{4\pi/L, -} \eta_{2\pi/L, -} \eta_{2\pi/L, -}^\dagger \eta_{4\pi/L, -}^\dagger...\eta_{2\pi-2\pi/L, -}^\dagger \eta_{0, -}\eta_{0, -}^\dagger \ket{0}\notag\\
	&=(-1)^{L-1}=e^{i\pi(L-1)}=e^{i\pi(L+1)}.
\end{align}
\begin{equation}
\Rightarrow \quad \braket{G(0)}{G(2\pi)}=e^{i\pi(L+1)},
\end{equation}
so that
\begin{equation}
\mathrm{Arg}\,\braket{G(0)}{G(2\pi)}=\pi(L+1).
\label{eq:overlap}
\end{equation}

Plugging Eqs.~(\ref{eq:int_term}) and (\ref{eq:overlap}) in Eq.~(\ref{eq:berry_general}), we find that the extensive terms $\pm\pi(L+1)$ cancel exactly to give
\begin{equation}
\gamma
=-\frac{1}{4\pi}
\!\int_{0}^{2\pi}\! d\phi
\!\int_{\phi/L}^{2\pi+\phi/L}\!
d\kappa\,\frac{d\theta(\kappa)}{d\kappa}.
\end{equation}

The inner integral depends only on the total change of $\theta(\kappa)$ over a $2\pi$ interval, and is therefore independent of $\phi$.  
Performing the $\phi$ integral gives the final result
\begin{equation}
\gamma
=-\frac{1}{2}
\int_{0}^{2\pi} d\kappa\,\frac{d\theta(\kappa)}{d\kappa},
\end{equation}
which is precisely the single-particle Zak phase of the occupied band.  
For the SSH model this yields
\begin{equation}
\gamma =
\begin{cases}
0, & v>w,\\[2mm]
\pi, & v<w.
\end{cases}
\end{equation}  
This establishes that the MBBP of the half-filled noninteracting SSH chain coincides with the conventional single-particle Zak phase.

\subsection{Entanglement spectrum in the noninteracting limit}

In this section, we derive the many-body entanglement spectrum (ES) \cite{Li2008, Pollmann2010, Turner2011} of a generic noninteracting fermionic ground state and apply it to the half-filled SSH model.
The derivation follows from an explicit Schmidt decomposition of the ground state. 

Consider a lattice system divided into two spatial regions $\mathcal{A}$ and $\mathcal{B}$.
The full fermionic Hilbert space factorizes as
\begin{align}
\mathcal{H}=\mathcal{H}_\mathcal{A} \otimes \mathcal{H}_\mathcal{B}.
\end{align}
Let ${c_i^\dagger}$ be the original fermionic creation operators, where the index $i$ runs over all sites of the system.
We separate them into operators acting on the two subsystems,
\begin{align}
\{c_i^\dagger\}=\{a_m^\dagger\} \cup \{b_n^\dagger\},
\end{align}
with $a_m^\dagger$ creating fermions in region $\mathcal{A}$ and $b_n^\dagger$ in region $\mathcal{B}$. Let there be total $L$ sites, $L_\mathcal{A}$ in region $\mathcal{A}$ and $L_\mathcal{B}$ in region $\mathcal{B}$, with $L=L_\mathcal{A}+L_\mathcal{B}$.

For a noninteracting Hamiltonian with $N$ particles, the many-body ground state is a Slater determinant of $N$ occupied single-particle orbitals,
\begin{align}
\ket{G}=d_1^\dagger d_2^\dagger...d_N^\dagger \ket{0}.
\end{align}
Each occupied orbital can be expanded in terms of the $\mathcal{A}$ and $\mathcal{B}$ operators as
\begin{align}
d_\alpha^\dagger=\sum_m U_{\alpha m} a_m^\dagger+\sum_n V_{\alpha n} b_n^\dagger, 
\end{align}
where $U$ and $V$ are matrices of dimensions $N\cross L_\mathcal{A}$ and $N\cross L_\mathcal{B}$ respectively, that encode the amplitudes of the single-particle wave functions on the two subsystems. Collecting all orbitals together, this can be written compactly as
\begin{align}
\textbf{d}^\dagger = 
\begin{pmatrix}
U & V
\end{pmatrix}\begin{pmatrix}
\textbf{a}^\dagger\\
\textbf{b}^\dagger
\end{pmatrix},
\end{align}
where the matrix $\begin{pmatrix}
U & V
\end{pmatrix}$ is composed by concatenating matrices $U$ and $V$, and has dimension $N\cross L$.
Because the orbitals corresponding to operators ${d_\alpha^\dagger}$ are orthonormal, the matrices satisfy
\begin{align}\label{uud_vvd}
U U^\dagger + V V^\dagger = \mathbf{I}_N.
\end{align}
The Slater determinant $\ket{G}$ is not in general written in a Schmidt-decomposed form with respect to the $\mathcal{A}|\mathcal{B}$ partition.
However, by performing unitary rotations among the occupied orbitals, one can bring it into such a form.

Let $X$ be a unitary matrix that diagonalizes $U U^\dagger$.
\begin{align}
X U U^\dagger X^\dagger=\Lambda,
\end{align}
where $\Lambda = \mathrm{diag}(\lambda_1,\lambda_2,\ldots,\lambda_N)$ with $0\le\lambda_\mu\le 1$, which follows from Eq.~\ref{uud_vvd} and the positive semi-definite nature of $U U^\dagger$. The same matrix $X$ also diagonalizes $VV^\dagger$ which is evident if we multiply Eq.~\ref{uud_vvd} from left and right by $X$ and $X^\dagger$, respectively. This gives the eigenvalues of $VV^\dagger$ as $(1-\lambda_1,1-\lambda_2,\ldots,1-\lambda_N)$ with $0\le(1-\lambda_\mu)\le 1$. 

We define the rotated orbitals as 
\begin{align}
\tilde{d}_\alpha^\dagger=\sum_m \tilde{U}_{\alpha m} a_m^\dagger+\sum_n \tilde{V}_{\alpha n} b_n^\dagger, 
\end{align}
where $\tilde{U}=XU$ and $\tilde{V}=XV$, so that $\tilde U \tilde U^\dagger=\Lambda$ and $\tilde V \tilde V^\dagger=\mathbb{I}-\Lambda$. Redefining $\tilde{U}_{\alpha m}=\sqrt{\lambda_\alpha}\tilde{w}_{\alpha m}$ and $\tilde{V}_{\alpha n}=\sqrt{1-\lambda_\alpha}\tilde{z}_{\alpha n}$, we get
\begin{align}
\tilde{d}_\alpha^\dagger=\sqrt{\lambda_\alpha}\tilde{f}_{\alpha \mathcal{A}}^\dagger+\sqrt{1-\lambda_\alpha}\tilde{f}_{\alpha \mathcal{B}}^\dagger, 
\end{align}
where $\tilde{f}_{\alpha \mathcal{A}}^\dagger=\sum_m \tilde{w}_{\alpha m}a_m^\dagger$ and $\tilde{f}_{\alpha \mathcal{B}}^\dagger=\sum_n \tilde{z}_{\alpha n}b_n^\dagger$. This decomposition clarifies the physical meaning of the eigenvalues.
For a particle in rotated orbital $\tilde d_\alpha^\dagger$, the coefficient $\lambda_\alpha$ represents the probability that the particle occupies subsystem $\mathcal{A}$, while $1-\lambda_\alpha$ is the probability of finding it in subsystem $\mathcal{B}$. Consequently: 
\begin{itemize}
\item $\lambda_\alpha=1$ corresponds to an orbital fully localized in $\mathcal{A}$,

\item $\lambda_\alpha=0$ corresponds to an orbital fully localized in $\mathcal{B}$,

\item $0<\lambda_\alpha<1$ describes an orbital shared between the two subsystems,

\item $\lambda_\alpha=\tfrac12$ corresponds to a maximally entangled mode with equal weight in $\mathcal{A}$ and $\mathcal{B}$.
\end{itemize}
Using the rotated orbitals, the many-body ground state can now be written in a form that factorizes into independent contributions from each orbital,
\begin{align}
\ket{G}
= \prod_{\alpha=1}^{N}
\left(
\sqrt{\lambda_\alpha}\tilde f_{\alpha \mathcal{A}}^\dagger
+\sqrt{1-\lambda_\alpha}\tilde f_{\alpha \mathcal{B}}^\dagger
\right)|0\rangle .
\end{align}
This expression is in Schmidt-decomposed form with respect to the bipartition $\mathcal{A}|\mathcal{B}$.
Expanding the product shows that the state is a superposition of configurations, in which each orbital $\alpha$ is occupied either in subsystem $\mathcal{A}$ or in subsystem $\mathcal{B}$, with probabilities determined by $\lambda_\alpha$ and $1-\lambda_\alpha$, respectively.

Tracing out subsystem $\mathcal{B}$, the reduced density matrix of region $\mathcal{A}$ becomes a direct product over the individual orbital sectors,
\begin{align}
\rho_\mathcal{A}
= \mathrm{Tr}_\mathcal{B}|G\rangle\langle G|
= \bigotimes_{\alpha=1}^{N}
\begin{pmatrix}
\lambda_\alpha & 0 \\
0 & 1-\lambda_\alpha
\end{pmatrix}=\bigotimes_{\alpha=1}^N \rho_{\alpha \mathcal{A}}, \quad \text{where} \quad \rho_{\alpha \mathcal{A}}=\begin{pmatrix}
\lambda_\alpha & 0 \\
0 & 1-\lambda_\alpha
\end{pmatrix}.
\end{align}
Thus each eigenvalue $\lambda_\alpha$ of $UU^\dagger$ generates two eigenvalues of $\rho_{\alpha \mathcal{A}}$, namely $\lambda_\alpha$ and $1-\lambda_\alpha$.
The full spectrum of $\rho_\mathcal{A}$ consists of all possible products:
\begin{align}
\Lambda_{{n_\alpha}}
= \prod_{\alpha}
\lambda_\alpha^{n_\alpha}
(1-\lambda_\alpha)^{1-n_\alpha},
\qquad n_\alpha =0,1 .
\end{align}
Thus, finding $\{\lambda_\alpha\}$ is enough to obtain the many body ES. We call the spectrum of $-\log{\rho_\mathcal{A}}$ as the many-body ES.
The correlation matrix restricted to subsystem $\mathcal{A}$ is given by,
\begin{align}
C_{mn}=\langle G| a_m^\dagger a_n|G\rangle = (U^\dagger U)_{mn}.
\end{align}
Since $UU^\dagger$ and $U^\dagger U$ share the same nonzero eigenvalues, the same set ${\lambda_\alpha}$ is obtained directly by diagonalizing the correlation matrix.
Therefore, diagonalizing the single-particle correlation matrix $C$ provides all information necessary to construct the complete many-body ES. It is important note that this great simplification occurs only because there are no interactions and the $\rho_\mathcal{A}$ factorizes \cite{Peschel2003}. All this breaks down when many-body interactions are present, and the many-body ES has to be computed using numerical techniques.

A particularly important case arises when $\lambda_\alpha = \tfrac12$.
For such an orbital, the corresponding block of $\rho_\mathcal{A}$ becomes proportional to the identity matrix,
\begin{align}
\frac{1}{2}\begin{pmatrix}
1 & 0\\
0 & 1
\end{pmatrix},
\end{align}
which contributes a factor of two degeneracy to every many-body entanglement level.
If the correlation matrix possesses $n$ eigenvalues equal to $1/2$, the total degeneracy of the many-body ES is $2^{n}$.\\
\subsection{Entanglement spectrum of the SSH model}

We now apply the general formalism above to the half-filled SSH chain.
For a bipartition of the system into two equal halves, we compute the correlation matrix restricted to subsystem $\mathcal{A}$. In the SSH chain, each unit cell contains two sublattices, 
$A$ and $B$.
When the system is bi-partitioned into regions 
$\mathcal{A}$ and $\mathcal{B}$ (spatial regions), every unit cell index 
$m$ within the subsystem actually carries an additional sublattice label.
Therefore, the restricted correlation matrix is a matrix of the form:
\begin{align}
C_{i\alpha,j\beta}=\bra{G}c_{i\alpha}^\dagger c_{j\beta}\ket{G}, \quad \alpha, \beta \in \{A, B\},
\end{align}
where $i$, $j$ label unit cells belonging to the chosen subsystem, i.e., spatial region $\mathcal{A}$. For each pair of unit cells 
$i$ and $j$ inside subsystem $\mathcal{A}$, the correlation matrix contains a $2\cross 2$ block
\begin{align}
C_{i,j}=
\begin{pmatrix}
<c_{iA}^\dagger c_{jA}> & <c_{iA}^\dagger c_{jB}>\\
<c_{iB}^\dagger c_{jA}> & <c_{iB}^\dagger c_{jB}>
\end{pmatrix}.
\end{align}
The full matrix $C$ is obtained by arranging these blocks for all $i$, $j$ inside the subsystem. For a subsystem containing $L_\mathcal{A}$ unit cells, $C$ is therefore a Hermitian matrix of dimension $2L_\mathcal{A} \times 2L_\mathcal{A}$.

For the noninteracting SSH ground state, these correlators can be written analytically using the occupied Bloch eigenstates. At half filling, only the lower band is occupied. Let
\begin{align}
\ket{u_-(k)}=\frac{1}{\sqrt{2}}\begin{pmatrix}
-e^{i\theta(k)}\\1
\end{pmatrix}
\end{align}
be the normalized eigenvector of the lower band of the SSH Hamiltonian, with $\theta(k)=\text{Arg} [v+we^{-ik}]$. Using translational invariance, the correlation function between two unit cells separated by $r=i-j$ is
\begin{align}
C_{AA}(r)=\frac{1}{2L}\sum_ke^{-ikr}, \quad C_{AB}(r)=\frac{1}{2L}\sum_ke^{-ikr}e^{-i\theta(k)},\notag\\
C_{BA}(r)=\frac{1}{2L}\sum_ke^{-ikr}e^{i\theta(k)}, \quad C_{BB}(r)=\frac{1}{2L}\sum_ke^{-ikr}.
\end{align}
\begin{figure}[t]
	\centering
	\includegraphics[width=\columnwidth]{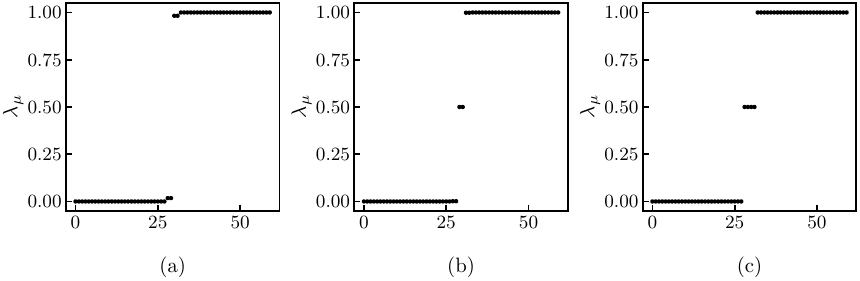}
	\caption{Eigenvalue spectra $\lambda_\mu$ of the restricted correlation matrix 
		$C$ for $L=60$.
		(a) Trivial phase $\gamma=0$ (or winding number $\nu=0$) : all eigenvalues lie near $0$ or $1$ and no eigenvalue is pinned at $1/2$.
		(b) Topological phase with $\gamma=\pi$ (or $\nu=1$): two eigenvalues are exactly $\lambda=1/2$, leading to a fourfold degeneracy of the many-body ES.
		(c) Higher-winding phase with $\nu=2$: four eigenvalues are pinned at $1/2$, producing a sixteenfold degeneracy of the ES.
		These spectra confirm that the number of eigenvalues pinned at $1/2$ equals $2\nu$.}
	\label{spectra_of_C}
\end{figure}
Here the Fourier convention is chosen such that $c_{j\alpha}=\frac{1}{\sqrt{L}}\sum_k e^{ijk}c_{k\alpha}$.
We diagonalize the restricted correlation matrix 
$C$ to obtain its eigenvalues $\{\lambda_\mu\}$, which completely determine the many-body ES. These eigenvalues are identical to those of $UU^\dagger$ obtained in the Schmidt decomposition, since $UU^\dagger$ and $U^\dagger U$ possess the same nonzero eigenvalues. This establishes the equivalence between the correlation-matrix approach and the explicit Schmidt construction.

Figs. \ref{spectra_of_C}(a)–(c) show the full eigenvalue spectra of $C$ for representative points in the trivial phase, the topological phase $\nu=1$, and the higher-winding phase $\nu=2$ of the extended model.

\begin{itemize}
\item Trivial phase ($v>w$, $\gamma=0 (\nu=0)$) — Fig. \ref{spectra_of_C}(a):
All eigenvalues of $C$ lie close to $0$ or $1$, with no eigenvalue pinned at $\lambda=1/2$.
Consequently the many-body ES is generically nondegenerate.

\item Topological SSH phase ($v<w$, $\gamma=\pi (\nu=1)$) — Fig. \ref{spectra_of_C}(b):
Two eigenvalues of $C$ are exactly pinned at $\lambda=1/2$.
Each such eigenvalue contributes a factor of two degeneracy to the many-body ES, resulting in a total fourfold degeneracy $2^2=4$, in agreement with the ES shown in the main text.

\item Higher-winding phase ($\nu=2$) — Fig. \ref{spectra_of_C}(c):
In the generalized SSH model with next-nearest-neighbor hopping, the restricted correlation matrix contains four eigenvalues pinned at $\lambda=1/2$.
These produce a sixteenfold degeneracy 
$2^4=16$ in the many-body ES.
\end{itemize}

These results demonstrate that the number of eigenvalues pinned at $1/2$ equals $2\nu$ in the phases considered here, where $\nu$ is the winding number.
Since each such eigenvalue contributes a factor of two to the degeneracy, the degeneracy of the many-body ES is therefore
\begin{align}
\text{ES degeneracy}=2^{2\nu}=4^\nu,
\end{align}
providing a quantitative realization of bulk–boundary correspondence in the ES. 

Connection to the ES in the main text—  
The analysis above yields the eigenvalues $\lambda_\alpha$ of the reduced density matrix $\rho_A$, obtained from the diagonalization of the restricted correlation matrix $C$.  
In the main text, however, the ES is displayed in terms of the eigenvalues of the entanglement Hamiltonian $H_E$, which is defined through
\begin{align}
	\rho_A = e^{-H_E}.
\end{align}
The two descriptions are fully equivalent and are related by a simple logarithmic transformation. Thus, the ES is actually the spectrum of $H_E$.

Given the single-particle eigenvalues $\{\lambda_\alpha\}$ of $C$, the reduced density matrix factorizes as a product of independent two-level sectors.  
Consequently, the entanglement Hamiltonian also has a free-fermion form \cite{Peschel2003},
\begin{align}
	H_E=\sum_{\alpha}\epsilon_\alpha\, 
	\tilde f_{\alpha A}^\dagger \tilde f_{\alpha A},
\end{align}
where the single-particle entanglement energies are
\begin{align}
	\epsilon_\alpha=\ln\!\left(\frac{1-\lambda_\alpha}{\lambda_\alpha}\right).
\end{align}

The many-body eigenvalues of $\rho_A$ are
\begin{align}
	\Lambda_{\{n_\alpha\}}
	=\prod_{\alpha}\lambda_\alpha^{n_\alpha}(1-\lambda_\alpha)^{1-n_\alpha},
	\qquad n_\alpha=0,1,
\end{align}
and the entanglement levels plotted in the main text are
\begin{align}
	\xi_{\{n_\alpha\}}=-\ln \Lambda_{\{n_\alpha\}}.
\end{align}

In particular, whenever a correlation-matrix eigenvalue satisfies $\lambda_\alpha=\tfrac12$, the corresponding entanglement energy is $\epsilon_\alpha=0$.  
Such zero modes of $H_E$ lead directly to degeneracies of the many-body ES: each eigenvalue $\lambda_\alpha=\tfrac12$ produces a twofold degeneracy, so that $n$ such eigenvalues generate a total degeneracy $2^{n}$.  
This mechanism explains the structures observed in Figs. 1(b), (c) of the main text and in Figs. \ref{spectra_of_C}(a–c) above:  
two eigenvalues pinned at $1/2$ yield a fourfold ES degeneracy in the SSH phase ($\nu=1$), while four such eigenvalues produce the sixteenfold degeneracy in the higher-winding phase ($\nu=2$).  
Thus the ES shown in the main text are simply the logarithmic representation of the same eigenvalues obtained here from the correlation-matrix approach.

\section{Mean-field description of interaction-induced parameter renormalization}

In this section, we provide a simple analytical understanding of the interaction-induced shift of the topological phase boundary discussed in the main text.  
Although the interacting model cannot be solved exactly, useful intuition can be obtained from a Hartree–Fock mean-field treatment, which captures how density–density interactions effectively renormalize the hopping amplitudes of the SSH chain \cite{Bruus2004}.

We consider the interacting Hamiltonian for periodic boundary condition:
\begin{align}
	\hat H_I &= \hat H_0
	+ \frac{U}{2}\sum_j \bigl(\hat{n}_{j,A}-\frac{1}{2}\bigr)\bigl(\hat{n}_{j,B}-\frac{1}{2}\bigr)
	+\frac{V}{2}\sum_j \bigl(\hat{n}_{j+1,A}-\frac{1}{2}\bigr)\bigl(\hat{n}_{j,B}-\frac{1}{2}\bigr)\notag\\
	&=\hat{H_0}+\frac{U}{2}\sum_j \hat{n}_{j,A}\hat{n}_{j,B}+\frac{V}{2}\sum_j \hat{n}_{j+1,A}\hat{n}_{j,B}-\frac{U+V}{4}\sum_j \bigl(\hat{n}_{j,A}+\hat{n}_{j,B}\bigr)+L\Bigl(\frac{U+V}{8}\Bigr) ,
\end{align}
where $\hat H_0$ is the noninteracting SSH Hamiltonian defined in the main text and $\hat{n}_{j,A(B)}=\hat{c}_{j,A(B)}^\dagger \hat{c}_{j,A(B)}$.  
To obtain a quadratic approximation, we decouple the interaction terms within the Hartree–Fock approximation.

For the intracell interaction, we write
\begin{align}
	\hat{n}_{j,A}\hat{n}_{j,B}
	&\approx \hat{n}_{j,A}\langle \hat{n}_{j,B}\rangle
	+\hat{n}_{j,B}\langle \hat{n}_{j,A}\rangle
	-\langle \hat{n}_{j,A}\rangle\langle \hat{n}_{j,B}\rangle \notag\\
	&\quad
	-\hat{c}_{j,A}^\dagger \hat{c}_{j,B}\langle \hat{c}_{j,B}^\dagger \hat{c}_{j,A}\rangle
	-\hat{c}_{j,B}^\dagger \hat{c}_{j,A}\langle \hat{c}_{j,A}^\dagger \hat{c}_{j,B}\rangle
	+|\langle \hat{c}_{j,B}^\dagger \hat{c}_{j,A}\rangle|^2 .
\end{align}

Assuming translational invariance in the bulk, the expectation values are taken to be independent of the unit cell index,
\begin{align}
	\langle \hat{n}_{j,A}\rangle=\bra{G}\hat{n}_{j,A}\ket{G}=n_A, \qquad
	\langle \hat{n}_{j,B}\rangle=\bra{G}\hat{n}_{j,B}\ket{G}=n_B, \qquad
	\langle \hat{c}_{j,B}^\dagger \hat{c}_{j,A}=\bra{G}\hat{c}_{j,B}^\dagger \hat{c}_{j,A}\rangle\ket{G}=\chi_U,
\end{align}
where $\ket{G}$ is the half-filled ground state of $\hat{H}_I$.
This gives
\begin{align}
	\hat{n}_{j,A}\hat{n}_{j,B}
	\approx \hat{n}_{j,A} n_{B}+\hat{n}_{j,B} n_{A}-n_{A} n_{B}
	-\hat{c}_{j,A}^\dagger \hat{c}_{j,B}\chi_U
	-\hat{c}_{j,B}^\dagger \hat{c}_{j,A}\chi_U^*
	+|\chi_U|^2 .
\end{align}

The intercell interaction is treated in the same way,
\begin{align}
	\hat{n}_{j+1,A}\hat{n}_{j,B}
	\approx \hat{n}_{j+1,A} n_{B}+\hat{n}_{j,B} n_{A}-n_{A} n_{B}
	-\hat{c}_{j+1,A}^\dagger \hat{c}_{j,B}\chi_V
	-\hat{c}_{j,B}^\dagger \hat{c}_{j+1,A}\chi_V^*
	+|\chi_V|^2 ,
\end{align}
with $\chi_V=\langle \hat{c}_{j,B}^\dagger \hat{c}_{j+1,A}\rangle=\bra{G}\hat{c}_{j,B}^\dagger \hat{c}_{j+1,A}\ket{G}$.

Substituting these expressions into $\hat H_I$ yields a quadratic mean-field Hamiltonian of SSH form,
\begin{align}\label{eq:H_MF}
	H_{\rm MF}
	&= \sum_j
	\left(
	v_{\rm eff}\, \hat{c}_{j,A}^\dagger \hat{c}_{j,B}
	+ w_{\rm eff}\, \hat{c}_{j+1,A}^\dagger \hat{c}_{j,B}
	+ {\rm H.c.}
	\right)  
	+ \sum_j (\mu_A n_{j,A}+\mu_B n_{j,B}) + L E_0 ,
\end{align}
with the following renormalized parameters:
\begin{align}\label{eq:MF_params}
	v_{\rm eff} &= v-\frac{U}{2}\chi_U, \qquad
	w_{\rm eff} = w-\frac{V}{2}\chi_V , \notag\\
	\mu_A &= \frac{U+V}{2} \bigl(n_B-\frac{1}{2}\bigr), \qquad
	\mu_B = \frac{U+V}{2} \bigl(n_A-\frac{1}{2}\bigr) , \notag\\
	E_0 &= -\frac{U+V}{2}\bigl(n_A n_B-\frac{1}{4}\bigr)
	+ \frac{U}{2}|\chi_U|^2
	+ \frac{V}{2}|\chi_V|^2 .
\end{align}

The effect of interactions is therefore twofold:

(i) the bare hoppings $v$ and $w$ are renormalized to effective values $v_{\rm eff}$ and $w_{\rm eff}$, and  

(ii) sublattice-dependent onsite potentials $\mu_A$ and $\mu_B$ are generated.

At half filling with inversion symmetry, the self-consistent solution satisfies $n_A=n_B$, so that $\mu_A=\mu_B$.  
At the Hartree–Fock level, the interacting problem reduces to an effective SSH model with renormalized parameters. This provides an intuitive explanation for the shift of the phase boundary observed in the exact many-body calculations presented in the main text.

To determine the mean-field parameters $(n_A,n_B,\chi_U,\chi_V)$ self-consistently, we diagonalize $H_{\rm MF}$ in momentum space.  
Fourier transforming Eq.~(\ref{eq:H_MF}) gives
\begin{align}
	H_{\rm MF}
	=\sum_k 
	\begin{pmatrix}
		c_{kA}^\dagger & c_{kB}^\dagger
	\end{pmatrix}
	h(k)
	\begin{pmatrix}
		c_{kA} \\ c_{kB}
	\end{pmatrix}
	+ L E_0 ,
\end{align}
where
\begin{align}
	h(k)=
	\begin{pmatrix}
		\mu_A & g(k)\\
		g^*(k) & \mu_B
	\end{pmatrix},
	\qquad
	g(k)=v_{\rm eff}+w_{\rm eff}e^{-ik}.
\end{align}

Let $(a_k,b_k)^T$ be the eigenstate of $h(k)$ with the lower-band eigenvalue
\begin{align}
	E_-(k)=\frac{\mu_A+\mu_B}{2}
	-\sqrt{\frac{(\mu_A-\mu_B)^2}{4}+|g(k)|^2}.
\end{align}
At half filling, the many-body ground state $\ket{G}$ is obtained by filling all lower-band states.  
The mean-field parameters are then given by
\begin{align}
	n_A &= \frac{1}{L}\sum_k |a_k|^2, \qquad
	n_B = \frac{1}{L}\sum_k |b_k|^2, \notag\\
	\chi_U &= \frac{1}{L}\sum_k a_k b_k^*, \qquad
	\chi_V = \frac{1}{L}\sum_k e^{ik} a_k b_k^* .
\end{align}

These transcendental equations are solved iteratively: starting from initial seed values for the parameters, the Hamiltonian is diagonalized, new expectation values are computed, and the procedure is repeated until convergence.

This mean-field picture provides a simple explanation for the numerical results of the main text.

\paragraph{Case $U=V$:}
For symmetric interactions, the self-consistent solutions give
\begin{align}
	n_A=n_B=1/2, \qquad \chi_U=\chi_V .
\end{align}
From Eq.~(\ref{eq:MF_params}), it then follows that
\begin{align}
	v_{\rm eff}=w_{\rm eff} \quad \text{whenever} \quad v=w.
\end{align}
Hence the bulk gap closes exactly at $v=w$, independent of interaction strength.  
This reproduces the behavior observed numerically in Fig.~2(a) of the main text: the topological transition remains pinned to $v=w$ when $U=V$.

\paragraph{Case $U\neq V$:}
When the two interaction strengths differ, the self-consistent values $\chi_U$ and $\chi_V$ are no longer equal.  
The effective hoppings become
\begin{align}
	v_{\rm eff}=v-\frac{U}{2}\chi_U , \qquad
	w_{\rm eff}=w-\frac{V}{2}\chi_V ,
\end{align}
so that the gap closes when $v_{\rm eff}=w_{\rm eff}$ rather than at $v=w$.  
Depending on whether $U>V$ or $U<V$, this condition is satisfied at values of $w$ larger or smaller than $v$, respectively.  
This explains the interaction-driven shift of the critical point observed in the exact diagonalization results of the main text. One more thing to note here is that the phase diagram and the phase boundaries no longer depend just on the ratios of $v$ and $w$ as the gap closing depends on their individual values and also on the values of interaction strengths $U$ and $V$. 

\begin{figure}[h]
	\centering
	\includegraphics[width=0.55\textwidth]{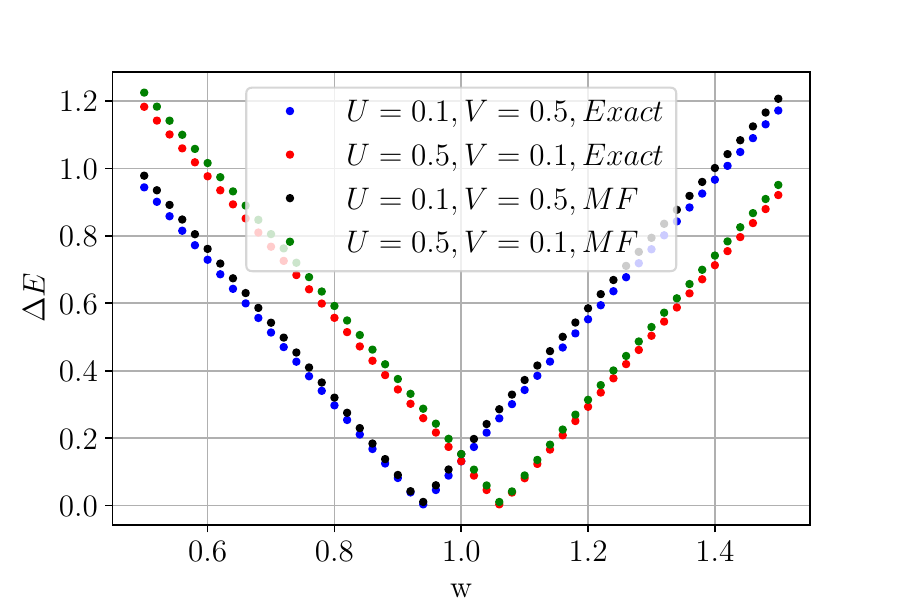}
	\caption{Comparison of the bulk excitation gap $\Delta E$ obtained from exact diagonalization for $L=10$ and that from the mean-field (MF) Hamiltonian when $U\neq V$ and $v=1$. Both methods show that the gap-closing point shifts to $w>v$ when $U>V$ and to $w<v$ when $U<V$, in agreement with the interacting phase diagram of Fig.~2(a) in the main text.}
	\label{fig:MF_gap}
\end{figure}

To illustrate this behavior, Fig.~\ref{fig:MF_gap} compares the excitation gap obtained from exact diagonalization with that predicted by the mean-field Hamiltonian for representative choices of $U$ and $V$.  
Even for modest system sizes, the mean-field treatment correctly captures the direction and approximate magnitude of the shift of the gap-closing point, demonstrating that the renormalization of the effective hoppings provides a reliable qualitative explanation of the interacting phase diagram.

We emphasize that this mean-field approach is intended only to provide qualitative intuition.  
At large interaction strengths, quantum fluctuations neglected in Hartree–Fock become important and the approximation may break down.  
Nevertheless, for the moderate interactions, the mean-field description captures the effective renormalization of hopping amplitudes and explains the observed movement of the phase boundary.

At the mean-field level, the interaction effects reduce to a renormalization of the hopping amplitudes and a uniform shift of the onsite potentials, such that the system is effectively described by a non-interacting SSH model with modified parameters.
Interestingly, this effective noninteracting description captures not only the shift of the phase boundary but also the many-body topological properties. In particular, the MBBP and the entanglement spectrum computed from the mean-field Hamiltonian are in good agreement with the exact diagonalization results. 

We note that related interaction-induced topological transitions have been studied in nonlinear frameworks based on self-consistent calculations, leading to nonlinear topological invariants such as the nonlinear Chern number \cite{Sone2024}. In contrast, the present approach is based on a fully many-body formulation of the ground state.
\section{Many-Body Winding Invariant}\label{Pancharatnam_WN}

We consider the family of Hamiltonians $\hat H(\phi)$ obtained by adiabatic
insertion of a boundary twist $\phi\in[0,2\pi]$. We assume that the many-body
ground state $|G(\phi)\rangle$ remains nondegenerate and separated from excited
states by a finite gap throughout the flux cycle. 

The MBBP associated with this evolution can be constructed from
overlaps of neighboring ground states along the path. In discrete form, it is given by
\begin{equation}
\gamma
=
\mathrm{Arg}\,
\prod_{j=0}^{N_\phi-1}
\langle G(\phi_j)|G(\phi_{j+1})\rangle,
\label{eq:berry_discrete_SM}
\end{equation}
with $\phi_{N_\phi} \equiv \phi_0$. This expression is manifestly gauge invariant,
since each phase factor appears together with its complex conjugate under a gauge
transformation. Because the Berry phase is defined through the argument,
it is only defined modulo $2\pi$. As a result, it characterizes the geometric
phase accumulated along the path but does not distinguish paths that differ by
additional windings in Hilbert space.

To construct an integer-valued invariant that resolves such higher windings,
we introduce a fixed reference state
\begin{equation}
|G_{\rm ref}\rangle \equiv |G(0)\rangle,
\end{equation}
and define the Bargmann invariant associated with each infinitesimal Pancharatnam triangular
loop formed by $|G(\phi_j)\rangle$, $|G(\phi_{j+1})\rangle$, and $|G_{\rm ref}\rangle$ \cite{Pancharatnam1956, Samuel1988}:
\begin{equation}
\Theta_j
=
\mathrm{Arg}\,
\left[
\langle G_{\rm ref}|G(\phi_j)\rangle\langle G(\phi_j)|G(\phi_{j+1})\rangle
\langle G(\phi_{j+1})|G_{\rm ref}\rangle\right].
\label{eq:bargmann_WN}
\end{equation}

The many-body winding invariant is then defined as
\begin{equation}
\nu
=
\frac{1}{\pi}
\sum_{j=0}^{N_\phi-1}
\Theta_j.
\label{eq:WN_discrete_SM}
\end{equation}

This construction explicitly incorporates the overlap with a fixed reference state
and therefore measures the winding of the ground state relative to a fixed reference state. We use this formula to construct the phase diagram in Fig. 3(a) of the main text, where we  numerically compute the winding invariant. Unlike the Berry phase, which is defined modulo $2\pi$,
the invariant $\nu$ resolves the full winding structure of the ground-state path.
In particular, $\nu$ captures phase discontinuities that occur when the overlap
$\langle G_{\rm ref}|G(\phi)\rangle$ vanishes. Such orthogonality points correspond
to singularities and contribute quantized $\pi$ increments
to $\nu$. As a result, $\nu$ takes integer values,
\begin{equation}
\nu \in \mathbb{Z},
\end{equation}
and defines a topological invariant of the interacting many-body ground state.

To obtain the continuous form of the invariant, we define the phase of the overlap
with the reference state,
\begin{equation}
\chi(\phi)
=
\mathrm{Arg}\,
\langle G(\phi)|G_{\rm ref}\rangle.
\label{eq:chi_definition_SM}
\end{equation}
Identifying $\Theta_j \equiv \Theta(\phi_j)$ and taking continuum limit, we can write $\Theta(\phi)$ in terms of $\chi$ as
\begin{align}\label{eq:bargmann_WN1}
\Theta(\phi)=\chi(\phi+\delta \phi)-\chi(\phi)+\mathrm{Arg}\, \langle G(\phi)|G(\phi+\delta \phi)\rangle.
\end{align}
For an infinitesimal increment $\delta\phi$, the overlap between neighboring
ground states satisfies
\begin{equation}
\langle G(\phi)|G(\phi+\delta\phi)\rangle
=
1
+
\delta\phi
\langle G(\phi)|\partial_\phi G(\phi)\rangle
+
\mathcal{O}(\delta\phi^2).
\end{equation}
Denoting $A(\phi)
=
i\langle G(\phi)|\partial_\phi G(\phi)\rangle$, we get
\begin{align}
\mathrm{Arg}\, \langle G(\phi)|G(\phi+\delta\phi)\rangle=\delta \phi A(\phi).\label{Arg1}
\end{align}
The remaining terms give
\begin{equation}
\chi(\phi+\delta \phi)-\chi(\phi)=\delta \phi \frac{d\chi}{d\phi}.\label{Arg2}
\end{equation}

Substituting Eqs.~\ref{Arg1} and \ref{Arg2} into Eq.~(\ref{eq:bargmann_WN1}) yields
\begin{equation}
\Theta(\phi)
=
\delta\phi
\left[
A(\phi)
+
\frac{d\chi}{d\phi}
\right].
\end{equation}

Summing over all intervals and taking the limit $N_\phi\to\infty$ give the continuous expression of Eq.~\ref{eq:WN_discrete_SM}:
\begin{equation}
\nu
=
\frac{1}{\pi}
\int_0^{2\pi}
d\phi
\left[
A(\phi)
+
\frac{d\chi}{d\phi}
\right],
\label{eq:WN_cont_SM}
\end{equation}
where $A(\phi)$ is the many-body Berry connection.

The gauge invariance of Eq.~(\ref{eq:WN_cont_SM}) can be verified explicitly.
Under a gauge transformation $|G(\phi)\rangle \rightarrow e^{i\alpha(\phi)}|G(\phi)\rangle$,
the Berry connection and overlap phase transform as
\begin{align}
A(\phi) &\rightarrow A(\phi) - \frac{d\alpha}{d\phi}, \\
\frac{d\chi}{d\phi} &\rightarrow \frac{d\chi}{d\phi} + \frac{d\alpha}{d\phi}.
\end{align}
The gauge-dependent terms cancel exactly, ensuring that $\nu$ is gauge invariant.

\section{Disorder implementation and symmetry protection of the many-body winding invariant}

In this section, we describe the implementation of disorder and its effect on the MBBP and the ES.

\subsection{Generic onsite disorder}

We introduce the onsite disorder in the Hamiltonian as:
\begin{align}
\hat{H}_{\mathrm{dis}}
=
\sum_j
\left(
\epsilon_{j,A} \hat{n}_{j,A}
+
\epsilon_{j,B} \hat{n}_{j,B}
\right),
\end{align}
where the onsite energies $\epsilon_{j,\alpha}$ are independently and identically drawn from a uniform distribution
\begin{align}
\epsilon_{j,\alpha} \in [-D, D],
\end{align}
with disorder strength $D$.

This form of disorder breaks translational and chiral symmetry of the model, in general, and also breaks inversion symmetry for each disorder realization.

\subsection{Inversion-symmetric disorder}

To isolate the role of inversion symmetry, we also consider disorder realizations that preserve inversion symmetry exactly.

For a chain of length $L$, inversion symmetry requires
\begin{align}
\epsilon_{j,A}
&=
\epsilon_{L+1-j,B}, \\
\epsilon_{j,B}
&=
\epsilon_{L+1-j,A}.
\end{align}

This constraint ensures that the full Hamiltonian remains invariant under the inversion operation
\begin{align}
\hat{c}_{j,A} \rightarrow \hat{c}_{L+1-j,B}, \qquad
\hat{c}_{j,B} \rightarrow \hat{c}_{L+1-j,A}.
\end{align}

Each inversion-symmetric disorder realization is constructed by generating random onsite potentials for half of the system and assigning the remaining values using the symmetry constraint above.

\subsection{Berry phase in the presence of disorder}

To quantify the effect of disorder on the topology, we monitor the MBBP of the interacting SSH model ($t=0$) accumulated under twisted boundary conditions as a function of the parameter $w$ with disorder.

For each disorder realization, we impose twisted boundary conditions by modifying the boundary hopping term as
\begin{align}
\hat{c}_{1,A}^\dagger \hat{c}_{L,B}
\;\rightarrow\;
e^{i\phi}
\hat{c}_{1,A}^\dagger \hat{c}_{L,B},
\end{align}
and compute the many-body ground state
\begin{align}
\ket{G(\phi)}
\end{align}
for a discrete set of twist angles
\begin{align}
\phi_m = \frac{2\pi m}{M}, \qquad m = 0, \dots, M-1.
\end{align}

To characterize the stability of the topological phase under disorder, we evaluate the Berry phase as a function of $w$ at a fixed disorder strength $D$.

For each value of $w$, we generate multiple disorder realizations and compute the corresponding many-body ground states and Berry phases.

The results shown in Fig.~4(a,b) of the main text are for two types of disorder:

\begin{enumerate}

\item Generic disorder, where onsite energies are chosen independently so that both translation and inversion symmetries are broken.

\item Inversion-symmetric disorder, where onsite energies satisfy the inversion constraint described earlier but the translation symmetry is broken.

\end{enumerate}

For generic disorder, the MBBP deviates continuously from its quantized value as disorder is introduced. This behavior reflects the absence of symmetry protection and indicates that the many-body topological invariant is no longer quantized when inversion symmetry is broken.

In contrast, for inversion-symmetric disorder, the MBBP remains quantized to $0$ and $\pi$ even with disorder. This demonstrates that inversion symmetry protects the quantization of the MBBP even in the presence of disorder breaking chiral and translational symmetry.

\subsection{Entanglement spectrum diagnostic in the presence of disorder}

To quantify the effect of disorder on the ES, we monitor the splitting between the lowest entanglement levels as a function of disorder strength.

For each disorder realization, we compute the many-body ground state under periodic boundary conditions and obtain the reduced density matrix $\rho_A$ by bipartitioning the system into two equal halves. The ES is defined as
\begin{align}
\xi_i = -\ln(\lambda_i),
\end{align}
where $\lambda_i$ are the eigenvalues of $\rho_A$, ordered as
\begin{align}
\xi_1 \le \xi_2 \le \xi_3 \le \cdots.
\end{align}

In the topological phase, the ES exhibits a characteristic fourfold degeneracy in the clean limit,
\begin{align}
\xi_1 = \xi_2 = \xi_3 = \xi_4.
\end{align}

To detect the lifting of this degeneracy under disorder, we define the entanglement splitting
\begin{align}
\Delta \xi = \xi_4 - \xi_1.
\end{align}

This quantity serves as a sensitive diagnostic of degeneracy lifting.

We evaluate $\Delta \xi$ as a function of disorder strength $D$ for a fixed parameter point deep in the topological phase. For each value of $D$, we generate multiple disorder realizations and compute the corresponding ground states and ES.

The results shown in Fig. 4(c,d) of the main text are for two types of disorder again:

\begin{enumerate}

\item Generic disorder, where onsite energies are chosen independently such that inversion symmetry is broken.

\item Inversion-symmetric disorder, where onsite energies satisfy the inversion constraint described in the previous subsection.

\end{enumerate}

For generic disorder, the splitting $\Delta \xi$ increases continuously with disorder strength, indicating lifting of the entanglement degeneracy and breakdown of the bulk–boundary correspondence.

In contrast, for inversion-symmetric disorder, the splitting remains exactly zero within numerical precision across the entire range of disorder strengths considered. This demonstrates that inversion symmetry protects the degeneracy structure of the ES even in the presence of strong disorder that breaks chiral symmetry.

These results provide a direct numerical confirmation that the many-body bulk–boundary correspondence established in the main text is protected by inversion symmetry even in the absence of chiral symmetry.

\subsection{Many-body winding invariant in the presence of disorder}

We now examine the behavior of the many-body winding invariant in the presence of disorder by varying the next-nearest-neighbor hopping parameter $t$ of $\hat{H}_{gen}$ while keeping all other parameters fixed as $v = 0.7, \quad w = 1, \quad U = V = 1, \quad D = 0.2$.

\begin{figure}[h]
	\centering
	\includegraphics[width=0.45\textwidth]{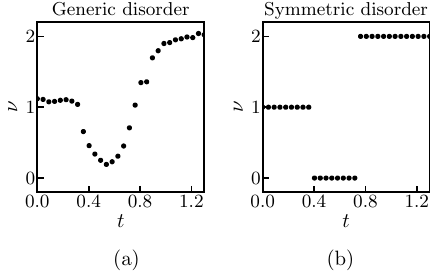}
	\caption{
Many-body winding invariant $\nu$ versus next-nearest-neighbor hopping $t$ for fixed $v=0.7$, $w=1$, $U=V=1$, and disorder strength $D=0.2$. The plots are generated for $L=6$. 
(a) Generic onsite disorder breaks inversion symmetry and destroys $\nu$'s quantization, resulting in smooth variation of $\nu$. It is averaged over 50 disorder realizations. 
(b) Inversion-symmetric disorder preserves quantization, with $\nu$ being locked to integer values and changing discontinuously at phase transitions.
}
	\label{fig:WN_dis_sym_dis}
\end{figure}

Figure~\ref{fig:WN_dis_sym_dis} shows the resulting winding invariant $\nu$ as a function of $t$:
Panel (a) corresponds to generic onsite disorder, while panel (b) corresponds to inversion-symmetric disorder.

In the case of generic disorder, the invariant $\nu$ is no longer quantized and varies smoothly as $t$ is tuned. The sharp transitions between distinct topological sectors observed in the clean limit are replaced by continuous crossovers, reflecting the absence of symmetry protection.

In contrast, when inversion symmetry is preserved, the winding invariant remains quantized to integer values within numerical precision. As $t$ crosses the topological phase boundaries, $\nu$ exhibits sharp jumps between integer values, signaling symmetry-protected topological phase transitions.

These results demonstrate that inversion symmetry alone is sufficient to protect the quantization of the many-body winding invariant even in the presence of disorder breaking translational and chiral symmetry.

\section{Onsite-modulated SSH Thouless pump}

Thus far we have focused on strictly one-dimensional (1D) lattice models. It is instructive, however, to consider a closely related system that effectively realizes higher-dimensional topological physics. To this end, we study an SSH chain with a periodically modulated onsite potential \cite{Li2014}. The Hamiltonian is given by
\begin{align}
\hat{H}_{T}=\hat{H}_{0}+\hat{H}_{\mu},
\end{align}
where 
\begin{align}
\hat{H}_{0}=\sum_j \Bigl(v\hat{c}_{j,A}^\dagger \hat{c}_{j,B}+w\hat{c}_{j+1,A}^\dagger \hat{c}_{j,B}+\text{H.c.}\Bigr),
\end{align}
and 
\begin{align}
\hat{H}_{\mu}=\sum_j \Bigl(\mu_A\hat{c}_{j,A}^\dagger \hat{c}_{j,A}+\mu_B\hat{c}_{j,B}^\dagger \hat{c}_{j,B}\Bigr).
\end{align}
The parameters are modulated as
\begin{align}
v&=1+\delta \cos \theta\notag\\
w&=1-\delta \cos \theta\notag\\
\mu_A&=g_A+h\cos(\theta-\tilde{\phi})\notag\\
\mu_B&=g_B+h\cos(\theta+\tilde{\phi}).
\end{align}

\begin{figure}[h]
	\centering
	\includegraphics[width=0.45\textwidth]{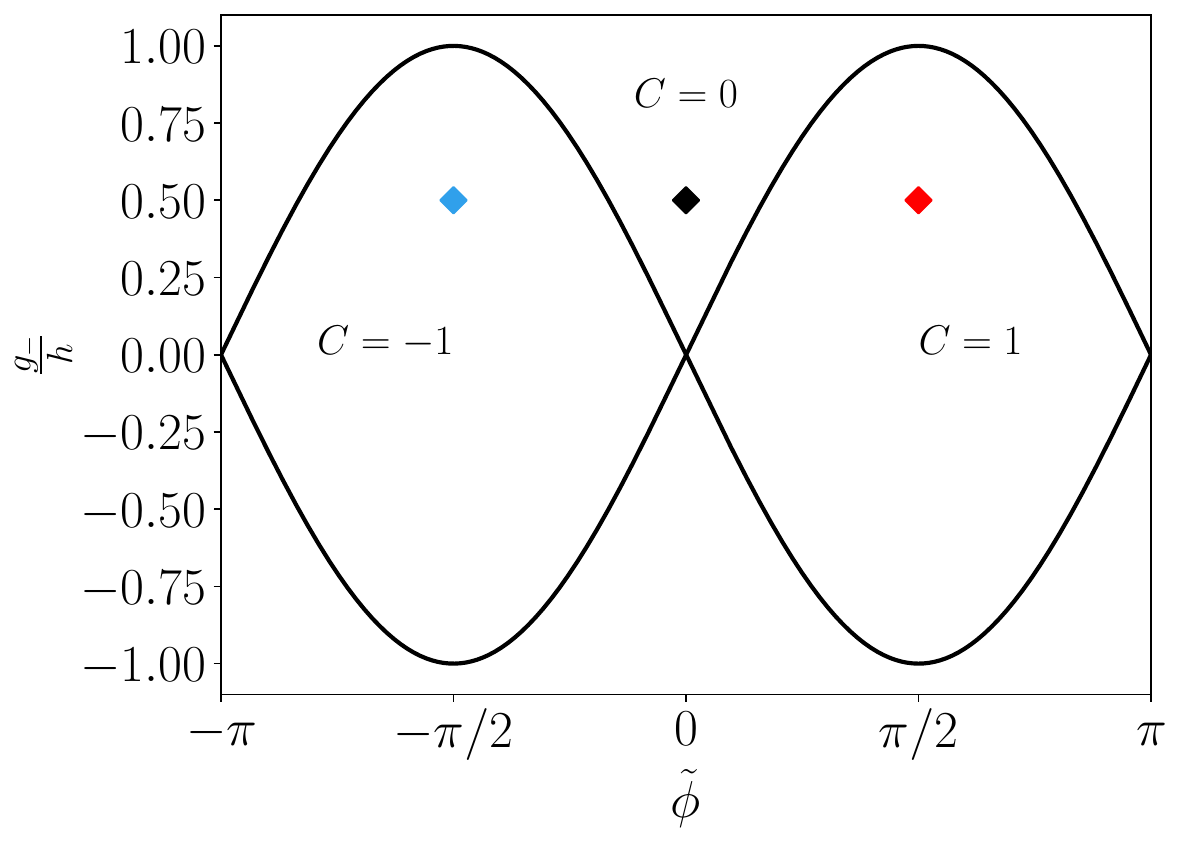}
	\caption{
Phase diagram of the onsite-modulated SSH model in the $(g_-/h,\tilde{\phi})$ plane, where $g_-=(g_A-g_B)/2$. The system exhibits topologically distinct phases characterized by Chern numbers $C=0,\pm1$. The phase boundaries correspond to gap-closing points of the bulk spectrum. For the range of interaction strengths considered in this work, the phase structure remains unchanged. Representative points used in the entanglement-spectrum analysis are indicated by blue, black and red diamonds.
}
	\label{fig:Mod_SSH_phase_diagram}
\end{figure}

The parameter $\theta$ can be interpreted as a synthetic momentum associated with an additional dimension, mapping the 1D model to an effective two-dimensional Chern insulator. This viewpoint provides a bridge between the 1D systems studied in the main text and higher-dimensional topological phases. The phase diagram, shown in Fig. \ref{fig:Mod_SSH_phase_diagram}, is characterized by regions with Chern numbers $C=0,\pm1$, and is controlled by the parameters $g_-/h$ and $\tilde{\phi}$, where $g_-=(g_A-g_B)/2$. For the interaction strengths considered below, the phase boundaries remain unchanged.

We now analyze the entanglement signatures of bulk–boundary correspondence in this model. For a bipartition of the chain into two equal spatial regions, we compute the eigenvalues $\{\lambda_\mu\}$ of the restricted correlation matrix as a function of the pump parameter $\theta$. As discussed in Sec.~\ref{ES_non_int}, the eigenvalues $\lambda_\mu=\tfrac12$ correspond to maximally entangled modes shared between the two subsystems and provide a direct entanglement signature of boundary degrees of freedom.

\begin{figure}[h]
	\centering
	\includegraphics[width=1\textwidth]{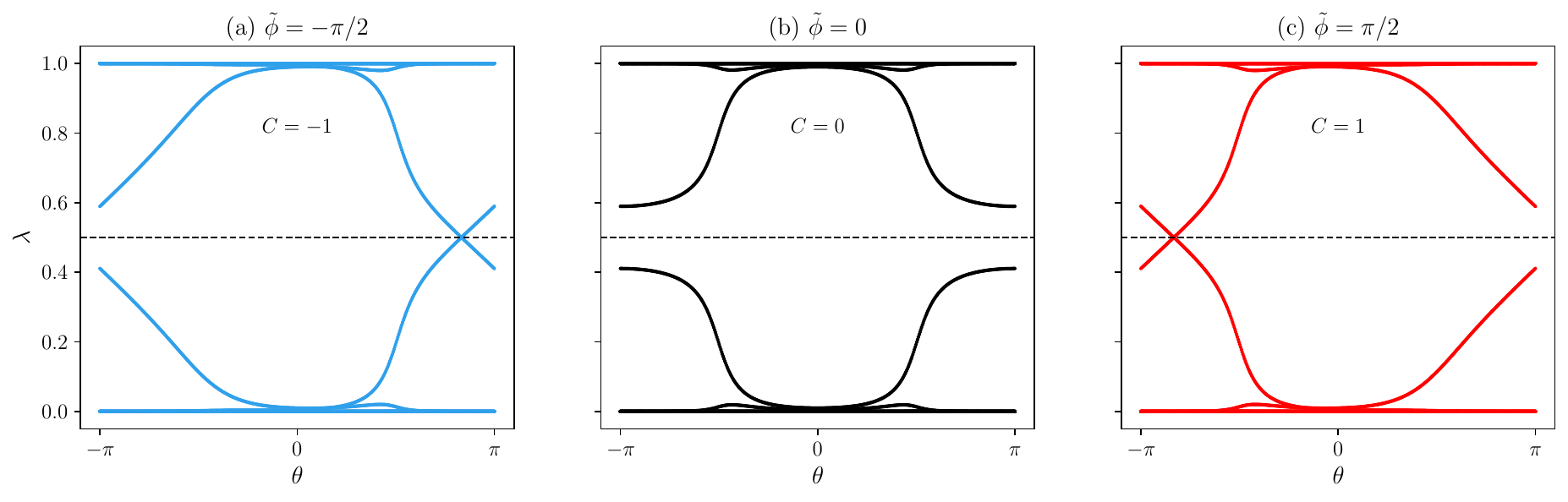}
	\caption{
Flow of the eigenvalues $\lambda_\mu$ of the restricted correlation matrix as a function of the pump parameter $\theta$ for the noninteracting many-body model. 
(b) Trivial phase, $C=0$ : all eigenvalues remain separated from $\lambda=\tfrac{1}{2}$ (dashed line) throughout the pump cycle. 
(a, c) Topological phase, $C=\pm 1$: the spectrum exhibits $\lambda=\tfrac{1}{2}$ crossings, signaling maximally entangled modes associated with boundary degrees of freedom. 
These crossings represent spectral flow in entanglement space and provide a signature of the underlying Chern topology. Parameters: $\delta=0.5$, $g_A=1$, $g_B=0.5$, $h=0.5$. All plots are generated for $L=20$.
}
	\label{fig:Non-int_mod_SSH_correlation_spec}
\end{figure}

We illustrate these features by selecting representative points from the trivial and topological regions of the phase diagram. Fig. \ref{fig:Non-int_mod_SSH_correlation_spec} shows the evolution of the correlation-matrix spectrum over one pump cycle. In the trivial phase ($C=0$), the eigenvalues remain separated from $\lambda=\tfrac12$ for all $\theta$. In contrast, in the topological phases ($C=\pm1$), the spectrum exhibits crossings at $\lambda=\tfrac12$. These crossings reflect the spectral flow associated with the underlying Chern topology and provide an entanglement manifestation of the bulk–boundary correspondence. 

To examine the interacting case, we consider the extended Hamiltonian $\hat{H}_{IT}= \hat H_T+\hat H_{\rm int}$ with
\begin{align}
	\hat{H}_\text{int} &= \frac{U}{2} \sum_j \Bigl(\Bigl(\hat{n}_{j, A}-\frac{1}{2}\Bigr) \Bigl(\hat{n}_{j, B}-\frac{1}{2}\Bigr)+ \Bigl(\hat{n}_{j+1, A}-\frac{1}{2}\Bigr) \Bigl(\hat{n}_{j, B}-\frac{1}{2}\Bigr)\Bigr),
\end{align}
which introduces nearest-neighbor density–density interactions. In the presence of interactions, the factorization of the reduced density matrix no longer holds, and the entanglement spectrum must be obtained directly from the many-body ground state.

\begin{figure}[h]
	\centering
	\includegraphics[width=1\textwidth]{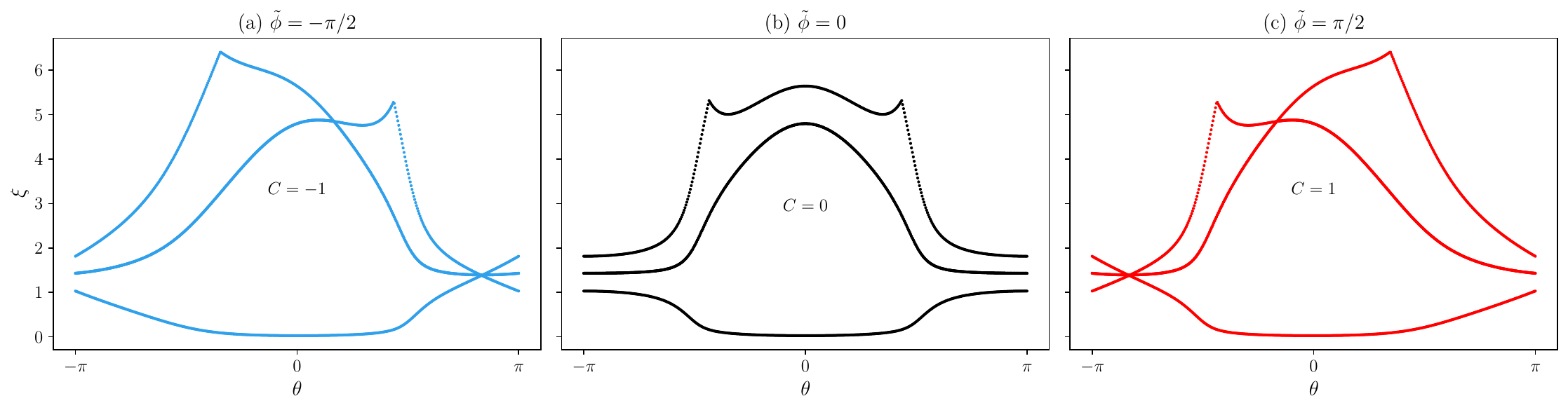}
	\caption{
Many-body entanglement spectrum, lowest four levels,  as a function of the pump parameter $\theta$ for the interacting model. 
(b) Trivial phase, $C=0$: the entanglement levels remain nondegenerate throughout the pump cycle. 
(a, c) Topological phase, $C=\pm 1$: the spectrum exhibits characteristic degeneracies, with levels forming fourfold multiplets. 
These degeneracies arise from maximally entangled boundary modes and represent the interacting counterpart of the $\lambda_\mu=\tfrac{1}{2}$ modes observed in the noninteracting correlation spectrum. 
The degeneracy structure provides a robust signature of the bulk–boundary correspondence that persists in the presence of interactions. Parameters: $\delta=0.5$, $g_A=1$, $g_B=0.5$, $h=0.5$, $U=0.5$. All plots are generated for $L=8$.
}
	\label{fig:Int_mod_SSH_ES_flow}
\end{figure}

Fig. \ref{fig:Int_mod_SSH_ES_flow} shows the lowest four levels of the resulting many-body entanglement spectrum as a function of $\theta$. In the trivial phase ($C=0$), the entanglement levels remain generically nondegenerate (with an accidental twofold degeneracy leading to three visible branches instead of four) throughout the pump cycle.  In contrast, in the topological phases ($C=\pm1$), the spectrum exhibits characteristic clustering of levels at specific values of $\theta$, where the levels become fourfold degenerate before separating again as $\theta$ is varied. This behavior reflects a nontrivial spectral flow and provides a signature of the underlying topology. These degeneracies represent the interacting counterpart of the $\lambda_\mu=\tfrac12$ modes observed in the correlation spectrum. More generally, the degeneracy structure follows the scaling $4^{|C|}$, consistent with the bulk–boundary correspondence established in the main text. If higher entanglement levels are included, additional clusters emerge in multiples of four, further validating the $4^{|C|}$ scaling.

Thus, while the correlation matrix provides a clear signature distinguishing trivial and nontrivial phases in the noninteracting limit, the full many-body entanglement spectrum continues to encode the bulk–boundary correspondence even in the presence of interactions.

\subsection{Relation between Chern number and the many-body winding invariant}
The Chern number for the interacting system can be computed using a discrete formulation of the Berry curvature by treating the pump parameter $\theta$ and twist angle $\phi$ as synthetic momenta $(k_x, k_y)$, following the approach of Fukui, Hatsugai and Suzuki \cite{Fukui2005}. In this method, the Berry curvature is evaluated on a discretized grid in the $(\theta, \phi)$ parameter space using gauge-invariant variables constructed from overlaps of many-body ground states. This provides a numerically stable and widely used procedure for computing the Chern number in interacting systems.

We now show that the many-body winding invariant defined in Sec.~\ref{Pancharatnam_WN} provides an alternative formulation that reduces the problem to a sequence of 1D flux-insertion cycles. In the onsite-modulated SSH model, the Chern number can be extracted from the winding of the Pancharatnam phase as a function of an external parameter.

For each value of $\theta$, we compute the many-body winding invariant $\nu(\theta)$ by performing a flux-insertion cycle over the twist angle $\phi\in[0,2\pi]$, as described in Sec.~\ref{Pancharatnam_WN}. This defines a parameter-dependent phase $\Phi(\theta)=\pi \nu(\theta)$ which captures the total Pancharatnam phase accumulated along the $\phi$-cycle.

The Chern number is then given by the winding of 
$\Phi(\theta)$ as $\theta$ is varied over a full cycle,
\begin{equation}
C = \frac{1}{2\pi} \int_0^{2\pi} d\theta  \frac{d\Phi(\theta)}{d\theta},
\end{equation}
which counts the number of times the many-body phase winds around the unit circle. This establishes a direct connection between the Chern number and the winding structure of the Pancharatnam phase constructed from the interacting ground state. Notably, while $\nu(\theta)$ is obtained from a 1D flux cycle in $\phi$, its evolution in $\theta$ encodes a two-dimensional topological invariant. 

\begin{figure}[h]
	\centering
	\includegraphics[width=0.45\textwidth]{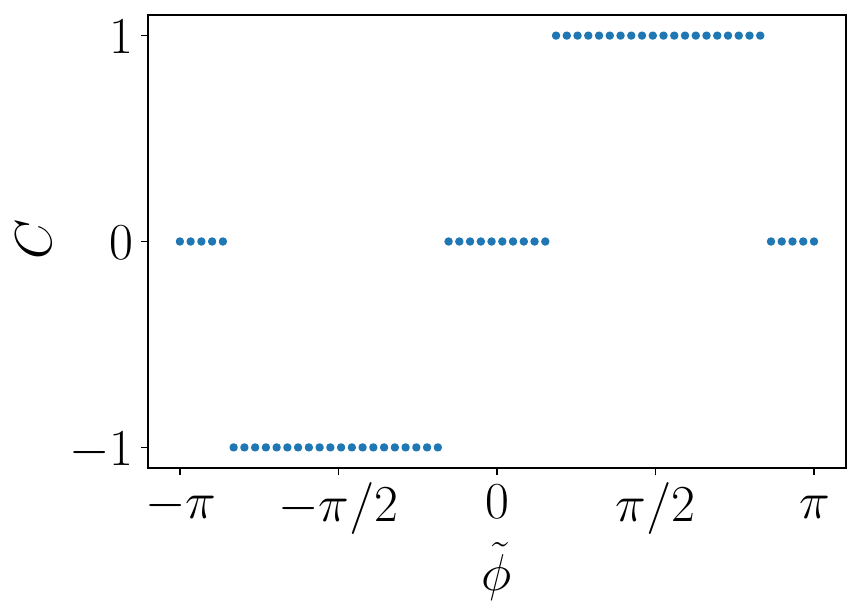}
	\caption{
Chern number $C$ as a function of the parameter $\theta$ for the interacting onsite-modulated SSH model for $L=6$, with parameters $\delta=0.5$, $g_A=1$, $g_B=0.5$, $h=0.5$, and $U=0.5$ (which gives $g_-/h=0.5$). The Chern number is computed using the many-body winding invariant. The quantized plateaus and transitions are in agreement with the phase boundaries shown in Fig.~\ref{fig:Mod_SSH_phase_diagram}, confirming the consistency of the two approaches. 
}
	\label{fig:Chern_no_int_mod_SSH}
\end{figure}

We plot the Chern number for the interacting onsite-modulated SSH model, calculated using this formulation, in Fig. \ref{fig:Chern_no_int_mod_SSH}. 

\makeatletter
\setcounter{NAT@ctr}{0}
\renewcommand{\theNAT@ctr}{\arabic{NAT@ctr}}
\makeatother

%% file: BBC_version_2.bbl
\begin{thebibliography}{0}%
\makeatletter
\providecommand \@ifxundefined [1]{%
 \@ifx{#1\undefined}
}%
\providecommand \@ifnum [1]{%
 \ifnum #1\expandafter \@firstoftwo
 \else \expandafter \@secondoftwo
 \fi
}%
\providecommand \@ifx [1]{%
 \ifx #1\expandafter \@firstoftwo
 \else \expandafter \@secondoftwo
 \fi
}%
\providecommand \natexlab [1]{#1}%
\providecommand \enquote  [1]{``#1''}%
\providecommand \bibnamefont  [1]{#1}%
\providecommand \bibfnamefont [1]{#1}%
\providecommand \citenamefont [1]{#1}%
\providecommand \href@noop [0]{\@secondoftwo}%
\providecommand \href [0]{\begingroup \@sanitize@url \@href}%
\providecommand \@href[1]{\@@startlink{#1}\@@href}%
\providecommand \@@href[1]{\endgroup#1\@@endlink}%
\providecommand \@sanitize@url [0]{\catcode `\\12\catcode `\$12\catcode
  `\&12\catcode `\#12\catcode `\^12\catcode `\_12\catcode `\%12\relax}%
\providecommand \@@startlink[1]{}%
\providecommand \@@endlink[0]{}%
\providecommand \url  [0]{\begingroup\@sanitize@url \@url }%
\providecommand \@url [1]{\endgroup\@href {#1}{\urlprefix }}%
\providecommand \urlprefix  [0]{URL }%
\providecommand \Eprint [0]{\href }%
\providecommand \doibase [0]{https://doi.org/}%
\providecommand \selectlanguage [0]{\@gobble}%
\providecommand \bibinfo  [0]{\@secondoftwo}%
\providecommand \bibfield  [0]{\@secondoftwo}%
\providecommand \translation [1]{[#1]}%
\providecommand \BibitemOpen [0]{}%
\providecommand \bibitemStop [0]{}%
\providecommand \bibitemNoStop [0]{.\EOS\space}%
\providecommand \EOS [0]{\spacefactor3000\relax}%
\providecommand \BibitemShut  [1]{\csname bibitem#1\endcsname}%
\let\auto@bib@innerbib\@empty
\end{thebibliography}%


\begin{thebibliography}{41}

\bibitem{Schnyder2008}
A. P. Schnyder, S. Ryu, A. Furusaki, and A. W. W. Ludwig,
Classification of topological insulators and superconductors in three spatial dimensions,
\href{https://link.aps.org/doi/10.1103/PhysRevB.78.195125}{Phys. Rev. B \textbf{78}, 195125 (2008)}.

\bibitem{Kitaev2009}
A. Kitaev, V. Lebedev, and M. Feigel’man,
Periodic table for topological insulators and superconductors,
\href{https://doi.org/10.1063/1.3149495}{AIP Conf. Proc., 22--30 (2009)}.

\bibitem{Thouless1982}
D. J. Thouless, M. Kohmoto, M. P. Nightingale, and M. den Nijs,
Quantized Hall Conductance in a Two-Dimensional Periodic Potential,
\href{https://link.aps.org/doi/10.1103/PhysRevLett.49.405}{Phys. Rev. Lett. \textbf{49}, 405 (1982)}.

\bibitem{Hatsugai1993}
Y. Hatsugai,
Chern number and edge states in the integer quantum Hall effect,
\href{https://link.aps.org/doi/10.1103/PhysRevLett.71.3697}{Phys. Rev. Lett. \textbf{71}, 3697 (1993)}.

\bibitem{Hasan2010}
M. Z. Hasan and C. L. Kane,
Colloquium: Topological insulators,
\href{https://link.aps.org/doi/10.1103/RevModPhys.82.3045}{Rev. Mod. Phys. \textbf{82}, 3045 (2010)}.

\bibitem{Goldman2016}
N. Goldman, J. C. Budich, and P. Zoller,
Topological quantum matter with ultracold gases in optical lattices,
\href{https://doi.org/10.1038/nphys3803}{Nat. Phys. \textbf{12}, 639 (2016)}.

\bibitem{Klembt2018}
S. Klembt, T. H. Harder, O. A. Egorov, K. Winkler, R. Ge, M. A. Bandres, M. Emmerling, L. Worschech, T. C. H. Liew, M. Segev, C. Schneider, and S. Höfling,
Exciton-polariton topological insulator,
\href{https://doi.org/10.1038/s41586-018-0601-5}{Nature \textbf{562}, 552 (2018)}.

\bibitem{Weber2022}
S. Weber, R. Bai, N. Makki, J. Mögerle, T. Lahaye, A. Browaeys, M. Daghofer, N. Lang, and H. P. Büchler,
Experimentally accessible scheme for a fractional Chern insulator in Rydberg atoms,
\href{https://link.aps.org/doi/10.1103/PRXQuantum.3.030302}{PRX Quantum \textbf{3}, 030302 (2022)}.

\bibitem{Alisepahi2023}
A. Rajabpoor Alisepahi, S. Sarkar, K. Sun, and J. Ma,
Breakdown of conventional winding number calculation in one-dimensional lattices with interactions beyond nearest neighbors,
\href{https://doi.org/10.1038/s42005-023-01461-0}{Commun. Phys. \textbf{6}, 334 (2023)}.

\bibitem{Sone2025}
K. Sone, M. Ezawa, Z. Gong, T. Sawada, N. Yoshioka, and T. Sagawa,
Transition from the topological to the chaotic in the nonlinear SSH model,
\href{https://doi.org/10.1038/s41467-024-55237-3}{Nat. Commun. \textbf{16}, 422 (2025)}.

\bibitem{Chen2013}
X. Chen, Z.-C. Gu, Z.-X. Liu, and X.-G. Wen,
Symmetry protected topological orders and the group cohomology of their symmetry group,
\href{https://link.aps.org/doi/10.1103/PhysRevB.87.155114}{Phys. Rev. B \textbf{87}, 155114 (2013)}.

\bibitem{David2022}
D. Mikhail, B. Voisin, D. D. St Medar, G. Buchs, S. Rogge, and S. Rachel,
Quasiparticle excitations in a one-dimensional interacting topological insulator,
\href{https://link.aps.org/doi/10.1103/PhysRevB.106.195408}{Phys. Rev. B \textbf{106}, 195408 (2022)}.

\bibitem{Li2023}
Y. Li, Y. Wang, H. Zhao, H. Du, J. Zhang, Y. Hu, F. Mei, L. Xiao, J. Ma, and S. Jia,
Interaction-induced breakdown of chiral dynamics in the SSH model,
\href{https://link.aps.org/doi/10.1103/PhysRevResearch.5.L032035}{Phys. Rev. Res. \textbf{5}, L032035 (2023)}.

\bibitem{Bisht2024}
J. Bisht, S. Jalal, and B. Kumar,
Transmigration of edge states with interaction in SSH chain,
\href{https://link.aps.org/doi/10.1103/PhysRevB.110.245110}{Phys. Rev. B \textbf{110}, 245110 (2024)}.

\bibitem{Fidkowski2010}
L. Fidkowski and A. Kitaev,
Effects of interactions on the topological classification of free fermion systems,
\href{https://link.aps.org/doi/10.1103/PhysRevB.81.134509}{Phys. Rev. B \textbf{81}, 134509 (2010)}.

\bibitem{Fidkowski2011}
L. Fidkowski and A. Kitaev,
Topological phases of fermions in one dimension,
\href{https://link.aps.org/doi/10.1103/PhysRevB.83.075103}{Phys. Rev. B \textbf{83}, 075103 (2011)}.

\bibitem{Turner2011}
A. M. Turner, F. Pollmann, and E. Berg,
Topological phases of one-dimensional fermions: An entanglement point of view,
\href{https://link.aps.org/doi/10.1103/PhysRevB.83.075102}{Phys. Rev. B \textbf{83}, 075102 (2011)}.

\bibitem{Gurarie2011}
V. Gurarie,
Single-particle Green's functions and interacting topological insulators,
\href{https://link.aps.org/doi/10.1103/PhysRevB.83.085426}{Phys. Rev. B \textbf{83}, 085426 (2011)}.

\bibitem{Niu1984}
Q. Niu and D. J. Thouless,
Quantised adiabatic charge transport in the presence of disorder and interactions,
\href{https://doi.org/10.1088/0305-4470/17/12/016}{J. Phys. A \textbf{17}, 2453 (1984)}.

\bibitem{Wang2010}
Z. Wang, X.-L. Qi, and S.-C. Zhang,
Topological order parameters for interacting topological insulators,
\href{https://link.aps.org/doi/10.1103/PhysRevLett.105.256803}{Phys. Rev. Lett. \textbf{105}, 256803 (2010)}.

\bibitem{Wang2012}
Z. Wang and S.-C. Zhang,
Simplified topological invariants for interacting insulators,
\href{https://link.aps.org/doi/10.1103/PhysRevX.2.031008}{Phys. Rev. X \textbf{2}, 031008 (2012)}.

\bibitem{Manmana2012}
S. R. Manmana, A. M. Essin, R. M. Noack, and V. Gurarie,
Topological invariants and interacting one-dimensional fermionic systems,
\href{https://link.aps.org/doi/10.1103/PhysRevB.86.205119}{Phys. Rev. B \textbf{86}, 205119 (2012)}.

\bibitem{Ke2017}
Y. Ke, X. Qin, Y. S. Kivshar, and C. Lee,
Multiparticle Wannier states and Thouless pumping of interacting bosons,
\href{https://link.aps.org/doi/10.1103/PhysRevA.95.063630}{Phys. Rev. A \textbf{95}, 063630 (2017)}.

\bibitem{Sone2024}
K. Sone, M. Ezawa, Y. Ashida, N. Yoshioka, and T. Sagawa,
Nonlinearity-induced topological phase transition characterized by the nonlinear Chern number,
\href{https://doi.org/10.1038/s41567-024-02451-x}{Nat. Phys. \textbf{20}, 1164 (2024)}.

\bibitem{Salvo2024}
E. Di Salvo, A. Moustaj, C. Xu, L. Fritz, A. K. Mitchell, C. Morais Smith, and D. Schuricht,
Topological phases of the interacting SSH model,
\href{https://link.aps.org/doi/10.1103/PhysRevB.110.165145}{Phys. Rev. B \textbf{110}, 165145 (2024)}.

\bibitem{Ortiz1994}
G. Ortiz and R. M. Martin,
Macroscopic polarization as a geometric quantum phase: Many-body formulation,
\href{https://link.aps.org/doi/10.1103/PhysRevB.49.14202}{Phys. Rev. B \textbf{49}, 14202 (1994)}.

\bibitem{Resta1998}
R. Resta,
Quantum-mechanical position operator in extended systems,
\href{https://link.aps.org/doi/10.1103/PhysRevLett.80.1800}{Phys. Rev. Lett. \textbf{80}, 1800 (1998)}.

\bibitem{Grusdt2019}
F. Grusdt, N. Y. Yao, and E. A. Demler,
Topological polarons, quasiparticle invariants, and their detection,
\href{https://link.aps.org/doi/10.1103/PhysRevB.100.075126}{Phys. Rev. B \textbf{100}, 075126 (2019)}.

\bibitem{Le2020}
N. H. Le, A. J. Fisher, N. J. Curson, and E. Ginossar,
Topological phases of a dimerized Fermi-Hubbard model,
\href{https://doi.org/10.1038/s41534-020-0253-9}{npj Quantum Inf. \textbf{6}, 24 (2020)}.

\bibitem{Li2008}
H. Li and F. D. M. Haldane,
Entanglement spectrum as a generalization of entanglement entropy,
\href{https://link.aps.org/doi/10.1103/PhysRevLett.101.010504}{Phys. Rev. Lett. \textbf{101}, 010504 (2008)}.

\bibitem{Pollmann2010}
F. Pollmann, A. M. Turner, E. Berg, and M. Oshikawa,
Entanglement spectrum of a topological phase in one dimension,
\href{https://link.aps.org/doi/10.1103/PhysRevB.81.064439}{Phys. Rev. B \textbf{81}, 064439 (2010)}.

\bibitem{Fidkowski2010(1)}
L. Fidkowski,
Entanglement spectrum of topological insulators and superconductors,
\href{https://link.aps.org/doi/10.1103/PhysRevLett.104.130502}{Phys. Rev. Lett. \textbf{104}, 130502 (2010)}.

\bibitem{Monkman2023}
K. Monkman and J. Sirker,
Entanglement and particle fluctuations of one-dimensional chiral topological insulators,
\href{https://link.aps.org/doi/10.1103/PhysRevB.108.125116}{Phys. Rev. B \textbf{108}, 125116 (2023)}.

\bibitem{Su1979}
W. P. Su, J. R. Schrieffer, and A. J. Heeger,
Solitons in polyacetylene,
\href{https://link.aps.org/doi/10.1103/PhysRevLett.42.1698}{Phys. Rev. Lett. \textbf{42}, 1698 (1979)}.

\bibitem{Pancharatnam1956}
S. Pancharatnam,
Generalized theory of interference, and its applications,
\href{https://doi.org/10.1007/BF03046050}{Proc. Indian Acad. Sci. A \textbf{44}, 247 (1956)}.

\bibitem{Samuel1988}
J. Samuel and R. Bhandari,
General setting for Berry's phase,
\href{https://link.aps.org/doi/10.1103/PhysRevLett.60.2339}{Phys. Rev. Lett. \textbf{60}, 2339 (1988)}.

\bibitem{SuppMat2026}
Supplementary Material.

\bibitem{Chen2020}
B.-H. Chen and D.-W. Chiou,
An elementary rigorous proof of bulk-boundary correspondence,
\href{https://doi.org/10.1016/j.physleta.2019.126168}{Phys. Lett. A \textbf{384}, 126168 (2020)}.

\bibitem{ES_Degeneracies_Comment}
In the noninteracting limit, this scaling can be derived analytically, explaining the $4^{\nu}$ degeneracy throughout the entanglement spectrum (see Supplemental Material).

\bibitem{Fuchs2021}
J.-N. Fuchs and F. Piéchon,
Orbital embedding and topology of one-dimensional two-band insulators,
\href{https://link.aps.org/doi/10.1103/PhysRevB.104.235428}{Phys. Rev. B \textbf{104}, 235428 (2021)}.

\bibitem{Li2014}
L. Li, Z. Xu, and S. Chen,
Topological phases of generalized SSH models,
\href{https://link.aps.org/doi/10.1103/PhysRevB.89.085111}{Phys. Rev. B \textbf{89}, 085111 (2014)}.

\end{thebibliography}

\begin{thebibliography}{99}

\bibitem{Su1979}
W. P. Su, J. R. Schrieffer, and A. J. Heeger,
Solitons in Polyacetylene,
\href{https://link.aps.org/doi/10.1103/PhysRevLett.42.1698}{Phys. Rev. Lett. \textbf{42}, 1698 (1979)}.

\bibitem{Ortiz1994}
G. Ortiz and R. M. Martin,
Macroscopic polarization as a geometric quantum phase: Many-body formulation,
\href{https://link.aps.org/doi/10.1103/PhysRevB.49.14202}{Phys. Rev. B \textbf{49}, 14202 (1994)}.

\bibitem{Resta1998}
R. Resta,
Quantum-Mechanical Position Operator in Extended Systems,
\href{https://link.aps.org/doi/10.1103/PhysRevLett.80.1800}{Phys. Rev. Lett. \textbf{80}, 1800 (1998)}.

\bibitem{Li2008}
Hui Li and F. D. M. Haldane,
Entanglement Spectrum as a Generalization of Entanglement Entropy: Identification of Topological Order in Non-Abelian Fractional Quantum Hall Effect States,
\href{https://link.aps.org/doi/10.1103/PhysRevLett.101.010504}{Phys. Rev. Lett. \textbf{101}, 010504 (2008)}.

\bibitem{Pollmann2010}
Frank Pollmann, Ari M. Turner, Erez Berg, and Masaki Oshikawa,
Entanglement spectrum of a topological phase in one dimension,
\href{https://link.aps.org/doi/10.1103/PhysRevB.81.064439}{Phys. Rev. B \textbf{81}, 064439 (2010)}.

\bibitem{Turner2011}
Ari M. Turner, Frank Pollmann, and Erez Berg,
Topological phases of one-dimensional fermions: An entanglement point of view,
\href{https://link.aps.org/doi/10.1103/PhysRevB.83.075102}{Phys. Rev. B \textbf{83}, 075102 (2011)}.

\bibitem{Peschel2003}
Ingo Peschel,
Calculation of reduced density matrices from correlation functions,
\href{https://doi.org/10.1088/0305-4470/36/14/101}{J. Phys. A \textbf{36}, L205 (2003)}.

\bibitem{Bruus2004}
Henrik Bruus and Karsten Flensberg,
Many-Body Quantum Theory in Condensed Matter Physics: An Introduction,
(Oxford University Press, 2004).

\bibitem{Sone2024}
Kazuki Sone, Motohiko Ezawa, Yuto Ashida, Nobuyuki Yoshioka, and Takahiro Sagawa,
Nonlinearity-induced topological phase transition characterized by the nonlinear Chern number,
\href{https://doi.org/10.1038/s41567-024-02451-x}{Nat. Phys. \textbf{20}, 1164 (2024)}.

\bibitem{Pancharatnam1956}
S. Pancharatnam,
Generalized theory of interference, and its applications,
\href{https://doi.org/10.1007/BF03046050}{Proc. Indian Acad. Sci. A \textbf{44}, 247 (1956)}.

\bibitem{Samuel1988}
Joseph Samuel and Rajendra Bhandari,
General Setting for Berry's Phase,
\href{https://link.aps.org/doi/10.1103/PhysRevLett.60.2339}{Phys. Rev. Lett. \textbf{60}, 2339 (1988)}.

\bibitem{Li2014}
Linhu Li, Zhihao Xu, and Shu Chen,
Topological phases of generalized Su-Schrieffer-Heeger models,
\href{https://link.aps.org/doi/10.1103/PhysRevB.89.085111}{Phys. Rev. B \textbf{89}, 085111 (2014)}.

\bibitem{Fukui2005}
Takahiro Fukui, Yasuhiro Hatsugai, and Hiroshi Suzuki,
Chern Numbers in Discretized Brillouin Zone: Efficient Method of Computing (Spin) Hall Conductances,
\href{https://doi.org/10.1143/JPSJ.74.1674}{J. Phys. Soc. Jpn. \textbf{74}, 1674 (2005)}.

\end{thebibliography}
